# Principal Fitted Components for Dimension Reduction in Regression

**R. Dennis Cook and Liliana Forzani**

*Abstract.* We provide a remedy for two concerns that have dogged the use of principal components in regression: (i) principal components are computed from the predictors alone and do not make apparent use of the response, and (ii) principal components are not invariant or equivariant under full rank linear transformation of the predictors. The development begins with principal fitted components [Cook, R. D. (2007). Fisher lecture: Dimension reduction in regression (with discussion). *Statist. Sci.* **22** 1–26] and uses normal models for the inverse regression of the predictors on the response to gain reductive information for the forward regression of interest. This approach includes methodology for testing hypotheses about the number of components and about conditional independencies among the predictors.

## 1. INTRODUCTION

Principal components have a long history of use as a dimension reduction method in regression, and today are widely represented in the applied sciences. The basic idea is to replace the predictor vector $\mathbf{X} \in \mathbb{R}^p$ with a few of the principal components $\hat{\mathbf{v}}_j^T \mathbf{X}$, $j = 1, \ldots, p$, prior to regression with response $Y \in \mathbb{R}^1$, where $\hat{\mathbf{v}}_j$ is the eigenvector of the sample covariance matrix $\hat{\Sigma}$ of $\mathbf{X}$ corresponding to its $j$th largest eigenvalue. The leading components, those corresponding to the larger eigenvalues, are typically chosen. Collinearity is the main and often the only motivation for the use of principal components in regression, but our results show no necessary link between the presence of collinearity and the effectiveness of principal component reduction.

Although collinearity is perhaps the primary historical reason for interest in principal components, they have been widely used in recent years for dimension reduction in regressions where the sample size $n$ is less than $p$. This motivation is prevalent in the genomics literature (see, e.g., Bura and Pfeiffer, 2003, and Li and Li, 2004) and is an important ingredient in the method of supervised principal components by Bair et al. (2006). Principal components can also be useful regardless of the presence of collinearity or the relationship between $n$ and $p$, depending on the goals of the analysis. For instance, it is often desirable to have an informative low-dimensional graphical representation of the data to facilitate model building and aid understanding (Cook, 1998). If $p \leq 2$ we can use computer graphics to view the data in full. If $p = 3$ and the response is categorical we can use a three-dimensional plot of the predictors with points marked according to the value of the response to view the full data. However, if $p > 3$ or $p = 3$ and the response is continuous we cannot view the data in full and dimension reduction may be useful.

*R. Dennis Cook is Professor, School of Statistics, University of Minnesota, 313 Ford Hall, 224 Church Street SE, Minneapolis, Minnesota 55455, USA e-mail: dennis@stat.umn.edu. Liliana Forzani is Professor, Facultad de Ingeniería Química, Universidad Nacional del Litoral and Instituto de Matemática Aplicada del Litoral, CONICET, Güemes 3450, (3000) Santa Fe, Argentina e-mail: liliana.forzani@gmail.com.*

Two general concerns have dogged the use of principal components. The first is that principal components are computed from the marginal distribution of $\mathbf{X}$ and there is no reason in principle why the leading principal components should carry the essential information about $Y$ (Cox, 1968). The second is that principal components are not invariant or equivariant under full rank linear transformations of $\mathbf{X}$, leading to problems in practice when the predictors are in different scales or have different variances. Some authors standardize the predictors so $\widehat{\boldsymbol{\Sigma}}$ is a correlation matrix prior to computing the components.

In this article we propose a model-based approach to principal component reduction that can be adapted to a specific response $Y$ and that is equivariant under full rank linear transformations of $\mathbf{X}$. Our results build on Cook's (2007) formulation, which uses model-based inverse regression of $\mathbf{X}$ on $Y$ to gain reductive information for the forward regression of $Y$ on $\mathbf{X}$. In Section 2 we introduce the models: Cook's models are reviewed in Section 2.1 and our extension is described in Section 2.2. We address estimation in Section 3 and relationships with other methods in Section 4. Inference is considered in Sections 5 and 6, and Section 7 contains illustrations of the proposed methodology. We regard the developments to this point as perhaps the most useful in practice. Nevertheless, in Section 8 we discuss model variations that may also be useful. Proofs can be found in the Appendices.

## 2. MODELS

We assume that the data consist of $n$ independent observations on $(\mathbf{X}, Y)$. Let $\mathbf{X}_y$ denote a random vector distributed as $\mathbf{X}|(Y = y)$, and assume that $\mathbf{X}_y$ is normally distributed with mean $\boldsymbol{\mu}_y$ and constant variance $\boldsymbol{\Delta} > 0$. Let $\bar{\boldsymbol{\mu}} = \mathrm{E}(\mathbf{X})$ and let $\mathcal{S}_{\boldsymbol{\Gamma}} = \mathrm{span}\{\boldsymbol{\mu}_y - \bar{\boldsymbol{\mu}}|y \in S_Y\}$, where $S_Y$ denotes the sample space of $Y$ and $\boldsymbol{\Gamma} \in \mathbb{R}^{p \times d}$ denotes a semi-orthogonal matrix whose columns form a basis for the $d$-dimensional subspace $\mathcal{S}_{\boldsymbol{\Gamma}}$. Then we can write

$$(1) \qquad \mathbf{X}_y = \bar{\boldsymbol{\mu}} + \boldsymbol{\Gamma}\boldsymbol{\nu}_y + \boldsymbol{\varepsilon},$$

where $\boldsymbol{\varepsilon}$ is independent of $Y$ and normally distributed with mean 0 and covariance matrix $\boldsymbol{\Delta}$, and $\boldsymbol{\nu}_y = \boldsymbol{\Gamma}^T(\boldsymbol{\mu}_y - \bar{\boldsymbol{\mu}}) \in \mathbb{R}^d$; we assume that $\mathrm{var}(\boldsymbol{\nu}_Y) > 0$. This model represents the fact that the translated conditional means $\boldsymbol{\mu}_y - \bar{\boldsymbol{\mu}}$ fall in the $d$-dimensional subspace $\mathcal{S}_{\boldsymbol{\Gamma}}$. In full generality, a reduction $T: \mathbb{R}^p \to$ $\mathbb{R}^q$, $q \leq p$, is sufficient if $Y|\mathbf{X} \sim Y|T(\mathbf{X})$, or equivalently if $Y \perp\!\!\!\perp \mathbf{X}|T(\mathbf{X})$, since $\mathbf{X}$ can then be replaced by $T(\mathbf{X})$ without loss of information on the regression. Under model (1) the specific reduction $R(\mathbf{X}) = \boldsymbol{\Gamma}^T\boldsymbol{\Delta}^{-1}\mathbf{X}$ is sufficient (Cook, 2007) and the goal is to estimate the *dimension reduction subspace* $\boldsymbol{\Delta}^{-1}\mathcal{S}_{\boldsymbol{\Gamma}} = \{\boldsymbol{\Delta}^{-1}\mathbf{z} : \mathbf{z} \in \mathcal{S}_{\boldsymbol{\Gamma}}\}$, since $\boldsymbol{\Gamma}$ is not generally identified. Then $R(\mathbf{X}) = \boldsymbol{\eta}^T\mathbf{X}$ is a sufficient reduction for any matrix $\boldsymbol{\eta} \in \mathbb{R}^{p \times d}$ whose columns form a basis for $\boldsymbol{\Delta}^{-1}\mathcal{S}_{\boldsymbol{\Gamma}}$. The parameter space for $\boldsymbol{\Delta}^{-1}\mathcal{S}_{\boldsymbol{\Gamma}}$ and $\mathcal{S}_{\boldsymbol{\Gamma}}$ is the $d$-dimensional Grassmann manifold $\mathcal{G}_{(d,p)}$ in $\mathbb{R}^p$. The manifold $\mathcal{G}_{(d,p)}$ has analytic dimension $d(p-d)$ (Chikuse, 2003, page 9), which is the number of reals needed to specify uniquely a single subspace in $\mathcal{G}_{(d,p)}$. This count will be used later when determining degrees of freedom.

Let $\mathcal{S}_d(\mathbf{A}, \mathbf{B})$ denote the span of $\mathbf{A}^{-1/2}$ times the first $d$ eigenvectors of $\mathbf{A}^{-1/2}\mathbf{B}\mathbf{A}^{-1/2}$, where $\mathbf{A}$ and $\mathbf{B}$ are symmetric matrices and, as used in this article, $\mathbf{A}$ will always be a nonsingular covariance matrix. Beginning with $\mathbf{B}$ we apply the transformation $\mathbf{A}^{-1/2}$ before computing the first $d$ eigenvectors. Multiplying these eigenvectors by $\mathbf{A}^{-1/2}$ then converts them to vectors that span the desired subspace in the original scale. The subspace $\mathcal{S}_d(\mathbf{A}, \mathbf{B})$ can also be described as the span of the first $d$ eigenvectors of $\mathbf{B}$ relative to $\mathbf{A}$. This notation is intended as a convenient way of describing estimators of $\boldsymbol{\Delta}^{-1}\mathcal{S}_{\boldsymbol{\Gamma}}$ under various conditions.

### 2.1 PC and Isotonic PFC Models

Cook (2007) developed estimation methods for two special cases of model (1). In the first, $\boldsymbol{\nu}_y$ is unknown for all $y \in S_Y$ but the errors are isotonic; that is, $\boldsymbol{\Delta} = \sigma^2\mathbf{I}_p$. This is called the *PC model* since the maximum likelihood estimator (MLE) of $\boldsymbol{\Delta}^{-1}\mathcal{S}_{\boldsymbol{\Gamma}} = \mathcal{S}_{\boldsymbol{\Gamma}}$ is $\mathcal{S}_d(\mathbf{I}_p, \widehat{\boldsymbol{\Sigma}})$, the span of the first $d$ eigenvectors of $\widehat{\boldsymbol{\Sigma}}$. Thus $R(\mathbf{X})$ is estimated by the first $d$ principal components. This relatively simple result arises because of the nature $\boldsymbol{\Delta}$. Since the errors are isotonic, the contours of $\boldsymbol{\Delta}$ are circular. When the signal $\boldsymbol{\Gamma}\boldsymbol{\nu}_y$ is added the contours of $\boldsymbol{\Sigma} = \boldsymbol{\Gamma}\,\mathrm{var}(\boldsymbol{\nu}_Y)\boldsymbol{\Gamma}^T + \sigma^2\mathbf{I}_p$ become $p$-dimensional ellipses with their longest $d$ axes spanning $\mathcal{S}_{\boldsymbol{\Gamma}}$.

In the second version of model (1), the coordinate vectors are modeled as $\boldsymbol{\nu}_y = \boldsymbol{\beta}\{\mathbf{f}_y - \mathrm{E}(\mathbf{f}_Y)\}$, where $\mathbf{f}_y \in \mathbb{R}^r$ is a known vector-valued function of $y$ with linearly independent elements and $\boldsymbol{\beta} \in \mathbb{R}^{d \times r}$, $d \leq \min(r, p)$, is an unrestricted rank $d$ matrix. Under this model for $\boldsymbol{\nu}_y$ each coordinate $X_{yj}$, $j = 1, \ldots, p$, of $\mathbf{X}_y$ follows a linear model with predictor vector $\mathbf{f}_y$. Consequently, we are able to use inverse response plots (Cook, [1998](#), Chapter 10) of $X_{yj}$ versus $y$, $j = 1, \ldots, p$, to gain information about suitable choices for $\mathbf{f}_y$, which is an ability that is not generally available in the forward regression of $Y$ on $\mathbf{X}$. For example, Bura and Cook ([2001](#)), Figure 1b, present a scatterplot matrix of the five variables in a regression with four predictors. The inverse response plots can all be fitted reasonably with $\log(y)$, indicating that in their example $\mathbf{f}_y = \log(y)$ may be adequate. In some regressions there may be a natural choice for $\mathbf{f}_y$. Suppose for instance that $Y$ is categorical, taking values in one of $h$ categories $C_k$, $k = 1, \ldots, h$. We can then set $r = h - 1$ and specify the $k$th element of $\mathbf{f}_y$ to be $J(y \in C_k)$, where $J$ is the indicator function. When $Y$ is continuous we can consider $\mathbf{f}_y$'s that contain a reasonably flexible set of basis functions, like polynomial terms in $Y$, which may be useful when it is impractical to apply graphical methods to all of the predictors. Another option consists of "slicing" the observed values of $Y$ into $h$ bins (categories) $C_k$, $k = 1, \ldots, h$, and then specifying the $k$th coordinate of $\mathbf{f}_y$ as for the case of a categorical $Y$. This has the effect of approximating each conditional mean $\mathrm{E}(X_{yj})$ as a step function of $y$ with $h$ steps,

$$\mathrm{E}(X_{yj}) \approx \bar{\mu}_j + \sum_{k=1}^{h-1} \gamma_j^T \mathbf{b}_k \{J(y \in C_k) - \Pr(Y \in C_k)\},$$

where $\gamma_j^T$ is the $j$th row of $\boldsymbol{\Gamma}$ and $\mathbf{b}_k$ is the $k$th column of $\boldsymbol{\beta}$. Piecewise polynomials could also be used.

Models with $\boldsymbol{\nu}_y = \boldsymbol{\beta}\{\mathbf{f}_y - \mathrm{E}(\mathbf{f}_Y)\}$ are called *principal fitted component (PFC) models*. When the errors are isotonic the MLE of $\mathcal{S}_{\boldsymbol{\Gamma}}$ is $\mathcal{S}_d(\mathbf{I}_p, \widehat{\boldsymbol{\Sigma}}_{\mathrm{fit}})$, where $\widehat{\boldsymbol{\Sigma}}_{\mathrm{fit}}$ is the sample covariance matrix of the fitted vectors from the multivariate linear regression of $\mathbf{X}_y$ on $\mathbf{f}_y$, including an intercept. In more detail, let $\mathbb{X}$ denote the $n \times p$ matrix with rows $(\mathbf{X}_y - \bar{\mathbf{X}})^T$ and let $\mathbb{F}$ denote the $n \times r$ matrix with rows $(\mathbf{f}_y - \bar{\mathbf{f}})^T$. Then the $n \times p$ matrix of fitted vectors from the regression of $\mathbf{X}_y$ on $\mathbf{f}_y$ is $\widehat{\mathbb{X}} = \mathbf{P}_{\mathbb{F}} \mathbb{X}$ and $\widehat{\boldsymbol{\Sigma}}_{\mathrm{fit}} = \mathbb{X}^T \mathbf{P}_{\mathbb{F}} \mathbb{X}/n$, where $\mathbf{P}_{\mathbb{F}}$ denotes the linear operator that projects onto the subspace spanned by the columns of $\mathbb{F}$. Under this *isotonic PFC model* the MLE of $R(\mathbf{X})$ consists of the first $d$ principal fitted components, with eigenvectors computed from $\widehat{\boldsymbol{\Sigma}}_{\mathrm{fit}}$ instead of $\widehat{\boldsymbol{\Sigma}} = \mathbb{X}^T \mathbb{X}/n$. The covariance matrix of the residual vectors from the fit of $\mathbf{X}_y$ on $\mathbf{f}_y$ can be represented as $\widehat{\boldsymbol{\Sigma}}_{\mathrm{res}} =$

$\widehat{\boldsymbol{\Sigma}} - \widehat{\boldsymbol{\Sigma}}_{\mathrm{fit}} = \mathbb{X}^T \mathbf{Q}_{\mathbb{F}} \mathbb{X}/n$, where $\mathbf{Q}_{\mathbb{F}} = \mathbf{I}_n - \mathbf{P}_{\mathbb{F}}$. This matrix plays no direct role in the isotonic PFC model, since we have specified $\boldsymbol{\Delta} = \sigma^2 \mathbf{I}_p$. However, $\widehat{\boldsymbol{\Sigma}}_{\mathrm{res}}$ will play a role in the extensions that follow.

## 2.2 The PFC Model

Principal fitted components are an adaptation of principal components to a particular response $Y$. However, the isotonic error structure $\boldsymbol{\Delta} = \sigma^2 \mathbf{I}_p$ is restrictive and does not address the invariance issue. In this article we extend principal fitted components to allow for a general error structure. Our specific goal is to develop maximum likelihood estimation of $\boldsymbol{\Delta}^{-1} \mathcal{S}_{\boldsymbol{\Gamma}}$ and related inference methods under the following PFC model,

$$(2) \quad \mathbf{X}_y = \bar{\boldsymbol{\mu}} + \boldsymbol{\Gamma}\boldsymbol{\beta}\{\mathbf{f}_y - \mathrm{E}(\mathbf{f}_Y)\} + \boldsymbol{\varepsilon} = \boldsymbol{\mu} + \boldsymbol{\Gamma}\boldsymbol{\beta}\mathbf{f}_y + \boldsymbol{\varepsilon},$$

where $\boldsymbol{\mu} = \bar{\boldsymbol{\mu}} - \boldsymbol{\Gamma}\boldsymbol{\beta}\mathrm{E}(\mathbf{f}_Y)$ and $\mathrm{var}(\boldsymbol{\varepsilon}) = \boldsymbol{\Delta} > 0$. This approach will then provide a solution to the two long-standing issues that have plagued the application of principal components. Assuming that $\widehat{\boldsymbol{\Sigma}}_{\mathrm{res}} > 0$, we will show that the MLE of the sufficient reduction $R(\mathbf{X})$ can be computed straightforwardly as the first $d$ principal components based on the standardized predictor vector $\widehat{\boldsymbol{\Sigma}}_{\mathrm{res}}^{-1/2} \mathbf{X}$. We found this result to be surprising since it does not depend explicitly on the MLE of $\boldsymbol{\Delta}$, which is a necessary ingredient for the MLE of $\boldsymbol{\Delta}^{-1} \mathcal{S}_{\boldsymbol{\Gamma}}$.

In Section [3](#) we give the MLE of $\boldsymbol{\Delta}$ and five equivalent forms for the MLE of $\boldsymbol{\Delta}^{-1} \mathcal{S}_{\boldsymbol{\Gamma}}$ under model ([2](#)). Relationships with some other methods are discussed in Section [4](#). We turn to inference in Sections [5](#) and [6](#), presenting ways of inferring about $d$ and about the active predictors. In Section [8](#) we discuss versions of model ([2](#)) in which $\boldsymbol{\Delta}$ is structured, providing modeling possibilities between the PFC models with $\boldsymbol{\Delta} = \sigma^2 \mathbf{I}_p$ and $\boldsymbol{\Delta} > 0$.

## 2.3 Identifying $\boldsymbol{\Delta}^{-1} \mathcal{S}_{\boldsymbol{\Gamma}}$ as the Central Subspace

In this section we provide a connection between the model-based dimension reduction considered in this article and the theory of model-free sufficient dimension reduction.

In the model-based approach, sufficient reductions $R(\mathbf{X})$ can in principle be determined from the model itself. For example, in the case of model ([1](#)) we saw previously that $R(\mathbf{X}) = \boldsymbol{\Gamma}^T \boldsymbol{\Delta}^{-1} \mathbf{X}$ is a sufficient reduction. In model-free dimension reduction there is no specific law to guide the choice of a reduction. However, progress is still possible by restricting attention to the class of linear reductions. A linear reduction always exists since $R(\mathbf{X}) = \mathbf{X}$ is trivially sufficient. If $R(\mathbf{X}) = \boldsymbol{\eta}^T\mathbf{X}$ is a $k$-dimensional sufficient reduction, then so is $R(\mathbf{X}) = (\boldsymbol{\eta}\mathbf{A})^T\mathbf{X}$ for any $k \times k$ full rank matrix $\mathbf{A}$. Consequently, only the subspace span($\boldsymbol{\eta}$) spanned by the columns of $\boldsymbol{\eta}$ can be identified —span($\boldsymbol{\eta}$) is called a *dimension reduction subspace*.

If span($\boldsymbol{\eta}$) is a dimension reduction subspace then so is span($\boldsymbol{\eta}, \boldsymbol{\eta}_1$) for any matrix $p \times k_1$ matrix $\boldsymbol{\eta}_1$. As a result there may be many linear sufficient reductions in a particular regression and we seek the one with the smallest dimension. If span($\boldsymbol{\eta}_1$) and span($\boldsymbol{\eta}_2$) are both dimension reduction subspaces, then under mild conditions (Cook, 1998) so is span($\boldsymbol{\eta}_1$) ∩ span($\boldsymbol{\eta}_2$). Consequently, the inferential target in model-free sufficient dimension reduction is often taken to be the central subspace $\mathcal{S}_{Y|\mathbf{X}}$, defined as the intersection of all dimension reduction subspaces (Cook, 1994, 1998).

The following theorem enables us to conclude that under model (2) $\boldsymbol{\Delta}^{-1}\mathcal{S}_{\boldsymbol{\Gamma}} = \mathcal{S}_{Y|\mathbf{X}}$; that is, the inferential targets for model-free and model-based reductions coincide in the context of model (2). The first part was given by Cook (2007), Proposition 6, but here we establish minimality as well.

THEOREM 2.1. *Let* $R(\mathbf{X}) = \boldsymbol{\Gamma}^T\boldsymbol{\Delta}^{-1}\mathbf{X}$, *and let* $T(\mathbf{X})$ *be any sufficient reduction. Then, under model (2),* $R$ *is a sufficient reduction and* $R$ *is a function of* $T$.

## 3. ESTIMATION UNDER PFC MODEL (2)

### 3.1 Maximum Likelihood Estimators

First we derive the MLE for the parameters of model (2) and then show how to linearly transform the predictors to yield a sufficient reduction. Our derivation requires that $\widehat{\boldsymbol{\Sigma}} > 0$, $\widehat{\boldsymbol{\Sigma}}_{\mathrm{res}} > 0$ and $d \leq \min(r, p)$. The full parameter space $(\boldsymbol{\mu}, \mathcal{S}_{\boldsymbol{\Gamma}}, \boldsymbol{\beta}, \boldsymbol{\Delta})$ for model (2) has analytic dimension $p + d(p - d) + dr + p(p+1)/2$. We hold $d$ fixed while maximizing over the other parameters, and then consider inference for $d$ in Section 5.

The full log likelihood is

$$L_d(\boldsymbol{\mu}, \mathcal{S}_{\boldsymbol{\Gamma}}, \boldsymbol{\beta}, \boldsymbol{\Delta})$$
$$= -\frac{np}{2}\log(2\pi) - (n/2)\log|\boldsymbol{\Delta}|$$
$$- (1/2)\sum_y (\mathbf{X}_y - \boldsymbol{\mu} - \boldsymbol{\Gamma}\boldsymbol{\beta}(\mathbf{f}_y - \bar{\mathbf{f}})^T)$$
$$\cdot \boldsymbol{\Delta}^{-1}(\mathbf{X}_y - \boldsymbol{\mu} - \boldsymbol{\Gamma}\boldsymbol{\beta}(\mathbf{f}_y - \bar{\mathbf{f}})),$$

(5)

where we have used the centered $\mathbf{f}_y$'s without loss of generality. For fixed $\boldsymbol{\Delta}$ and $\boldsymbol{\Gamma}$, this log likelihood is maximized over $\boldsymbol{\mu}$ and $\boldsymbol{\beta}$ by the arguments $\hat{\boldsymbol{\mu}} = \bar{\mathbf{X}}$ and $\tilde{\boldsymbol{\beta}} = \boldsymbol{\Gamma}^T\mathbf{P}_{\boldsymbol{\Gamma}(\boldsymbol{\Delta}^{-1})}\widehat{\mathbf{B}}$, where $\mathbf{P}_{\boldsymbol{\Gamma}(\boldsymbol{\Delta}^{-1})} = \boldsymbol{\Gamma}(\boldsymbol{\Gamma}^T\boldsymbol{\Delta}^{-1}\boldsymbol{\Gamma})^{-1} \cdot \boldsymbol{\Gamma}^T\boldsymbol{\Delta}^{-1}$ is the projection onto $\mathcal{S}_{\boldsymbol{\Gamma}}$ in the $\boldsymbol{\Delta}^{-1}$ inner product and $\widehat{\mathbf{B}} = \mathbb{X}^T\mathbb{F}(\mathbb{F}^T\mathbb{F})^{-1}$ is the coefficient matrix from the multivariate linear regression of $\mathbf{X}_y$ on $\mathbf{f}_y$. From the form of $\tilde{\boldsymbol{\beta}}$ we see that the MLE of $\boldsymbol{\Gamma}\boldsymbol{\beta}$ will be the projection of $\widehat{\mathbf{B}}$ onto $\hat{\mathcal{S}}_{\boldsymbol{\Gamma}}$ in the $\widehat{\boldsymbol{\Delta}}^{-1}$ inner product. To find $\hat{\mathcal{S}}_{\boldsymbol{\Gamma}}$ and $\widehat{\boldsymbol{\Delta}}$ we substitute $\hat{\boldsymbol{\mu}}$ and $\tilde{\boldsymbol{\beta}}$ into the log likelihood to obtain the partially maximized form

$$L_d(\mathcal{S}_{\boldsymbol{\Gamma}}, \boldsymbol{\Delta})$$
$$= -\frac{np}{2}\log(2\pi) - (n/2)\log|\boldsymbol{\Delta}|$$
$$- (n/2)\,\mathrm{trace}\{\boldsymbol{\Delta}^{-1/2}\widehat{\boldsymbol{\Sigma}}\boldsymbol{\Delta}^{-1/2}$$
$$- \mathbf{P}_{\boldsymbol{\Delta}^{-1/2}\boldsymbol{\Gamma}}\boldsymbol{\Delta}^{-1/2}\widehat{\boldsymbol{\Sigma}}_{\mathrm{fit}}\boldsymbol{\Delta}^{-1/2}\}.$$

(3)

Holding $\boldsymbol{\Delta}$ fixed, this is maximized by choosing $\mathbf{P}_{\boldsymbol{\Delta}^{-1/2}\boldsymbol{\Gamma}}$ to project onto the space spanned by the first $d$ eigenvectors of $\boldsymbol{\Delta}^{-1/2}\widehat{\boldsymbol{\Sigma}}_{\mathrm{fit}}\boldsymbol{\Delta}^{-1/2}$. This leads to the final partially maximized log likelihood (Cook, 2007, Section 7.2)

$$L_d(\boldsymbol{\Delta}) = -\frac{np}{2}\log(2\pi) - \frac{n}{2}\log|\boldsymbol{\Delta}|$$
$$- \frac{n}{2}\mathrm{tr}(\boldsymbol{\Delta}^{-1}\widehat{\boldsymbol{\Sigma}}_{\mathrm{res}}) - \frac{n}{2}\sum_{i=d+1}^{p}\lambda_i(\boldsymbol{\Delta}^{-1}\widehat{\boldsymbol{\Sigma}}_{\mathrm{fit}}),$$

(4)

where $\lambda_i(\mathbf{A})$ indicates the $i$th eigenvalue of the matrix $\mathbf{A}$. The MLE $\widehat{\boldsymbol{\Delta}}$ of $\boldsymbol{\Delta}$ is then obtained by maximizing (4). The MLEs of the remaining parameters are $\hat{\boldsymbol{\mu}} = \bar{\mathbf{X}}$, $\hat{\mathcal{S}}_{\boldsymbol{\Gamma}} = \widehat{\boldsymbol{\Delta}}\mathcal{S}_d(\widehat{\boldsymbol{\Delta}}, \widehat{\boldsymbol{\Sigma}}_{\mathrm{fit}})$, and $\hat{\boldsymbol{\beta}} = (\widehat{\boldsymbol{\Gamma}}^T\widehat{\boldsymbol{\Delta}}^{-1}\widehat{\boldsymbol{\Gamma}})^{-1} \cdot \widehat{\boldsymbol{\Gamma}}^{-T}\widehat{\boldsymbol{\Delta}}^{-1}\widehat{\mathbf{B}}$, where $\widehat{\boldsymbol{\Gamma}}$ is any orthonormal basis for $\hat{\mathcal{S}}_{\boldsymbol{\Gamma}}$. It follows that the sufficient reduction is of the form $\widehat{R}(\mathbf{X}) = (\widehat{\mathbf{v}}_1^T\widehat{\boldsymbol{\Delta}}^{-1/2}\mathbf{X}, \ldots, \widehat{\mathbf{v}}_d^T\widehat{\boldsymbol{\Delta}}^{-1/2}\mathbf{X})^T$, where $\widehat{\mathbf{v}}_j$ is the $j$th eigenvector of $\widehat{\boldsymbol{\Delta}}^{-1/2}\widehat{\boldsymbol{\Sigma}}_{\mathrm{fit}}\widehat{\boldsymbol{\Delta}}^{-1/2}$. The following theorem shows how to construct $\widehat{\boldsymbol{\Delta}}$.

THEOREM 3.1. *Let* $\widehat{\mathbf{V}}$ *and* $\widehat{\boldsymbol{\Lambda}} = \mathrm{diag}(\widehat{\lambda}_1, \ldots, \widehat{\lambda}_p)$ *be the matrices of the ordered eigenvectors and eigenvalues of* $\widehat{\boldsymbol{\Sigma}}_{\mathrm{res}}^{-1/2}\widehat{\boldsymbol{\Sigma}}_{\mathrm{fit}}\widehat{\boldsymbol{\Sigma}}_{\mathrm{res}}^{-1/2}$, *and assume that the nonzero* $\widehat{\lambda}_i$'s *are distinct. Then, the maximum of* $L_d(\boldsymbol{\Delta})$ *(4) over* $\boldsymbol{\Delta} > 0$ *is attained at* $\widehat{\boldsymbol{\Delta}} = \widehat{\boldsymbol{\Sigma}}_{\mathrm{res}} + \widehat{\boldsymbol{\Sigma}}_{\mathrm{res}}^{1/2}\widehat{\mathbf{V}}\widehat{\mathbf{K}}\widehat{\mathbf{V}}^T\widehat{\boldsymbol{\Sigma}}_{\mathrm{res}}^{1/2}$, *where* $\widehat{\mathbf{K}} = \mathrm{diag}(0, \ldots, 0, \widehat{\lambda}_{d+1}, \ldots, \widehat{\lambda}_p)$. *The maximum value of the log likelihood is*

$$L_d = -\frac{np}{2} - \frac{np}{2}\log(2\pi)$$

(5)

$$-\frac{n}{2}\log|\widehat{\boldsymbol{\Sigma}}_{\mathrm{res}}| - \frac{n}{2}\sum_{i=d+1}^{p}\log(1+\widehat{\lambda}_i).$$

In this theorem, $\widehat{\lambda}_i = 0$ for $i = r+1, \ldots, p$. Consequently, if $r = d$ then $\widehat{\boldsymbol{\Delta}} = \widehat{\boldsymbol{\Sigma}}_{\mathrm{res}}$, and the last term of $L_d$ does not appear. The maximum value of the log likelihood can also be expressed in terms of the squared sample canonical correlations $r_j^2$, $j = 1, \ldots, \min(p, r)$, between $\mathbf{X}$ and $\mathbf{f}_y$:

COROLLARY 3.2.

$$L_d = -\frac{np}{2} - \frac{np}{2}\log(2\pi)$$
$$- \frac{n}{2}\log|\widehat{\boldsymbol{\Sigma}}_{\mathrm{res}}| + \frac{n}{2}\sum_{i=d+1}^{\min(p,r)}\log(1-r_i^2).$$

The following corollary confirms the invariance of $\widehat{R}$ under full rank linear transformations of $\mathbf{X}$.

COROLLARY 3.3. *If $\mathbf{A} \in \mathbb{R}^{p \times p}$ has full rank, then $\widehat{R}(\mathbf{X}) = \widehat{R}(\mathbf{A}\mathbf{X})$.*

The next corollary gives five equivalent forms for the MLE of $\boldsymbol{\Delta}^{-1}\mathcal{S}_{\boldsymbol{\Gamma}}$.

COROLLARY 3.4. *The following are equivalent expressions for the MLE of $\boldsymbol{\Delta}^{-1}\mathcal{S}_{\boldsymbol{\Gamma}}$ under model (2): $\mathcal{S}_d(\widehat{\boldsymbol{\Delta}}, \widehat{\boldsymbol{\Sigma}}) = \mathcal{S}_d(\widehat{\boldsymbol{\Sigma}}_{\mathrm{res}}, \widehat{\boldsymbol{\Sigma}}) = \mathcal{S}_d(\widehat{\boldsymbol{\Delta}}, \widehat{\boldsymbol{\Sigma}}_{\mathrm{fit}}) = \mathcal{S}_d(\widehat{\boldsymbol{\Sigma}}_{\mathrm{res}}, \widehat{\boldsymbol{\Sigma}}_{\mathrm{fit}}) = \mathcal{S}_d(\widehat{\boldsymbol{\Sigma}}, \widehat{\boldsymbol{\Sigma}}_{\mathrm{fit}})$.*

The first and second forms—$\mathcal{S}_d(\widehat{\boldsymbol{\Delta}}, \widehat{\boldsymbol{\Sigma}}) = \mathcal{S}_d \times (\widehat{\boldsymbol{\Sigma}}_{\mathrm{res}}, \widehat{\boldsymbol{\Sigma}})$—indicate that the sufficient reduction under model (2) can be computed as the principal components based on the linearly transformed predictors $\widehat{\boldsymbol{\Delta}}^{-1/2}\mathbf{X}$ or $\widehat{\boldsymbol{\Sigma}}_{\mathrm{res}}^{-1}\mathbf{X}$, as mentioned in the Introduction. The remaining forms indicate that a sufficient reduction can be computed also as the principal fitted components for $\mathbf{A}^{-1/2}\mathbf{X}$, where $\mathbf{A}$ is $\widehat{\boldsymbol{\Delta}}$, $\widehat{\boldsymbol{\Sigma}}_{\mathrm{res}}$ or $\widehat{\boldsymbol{\Sigma}}$. Although $\widehat{\boldsymbol{\Delta}}$ and $\widehat{\boldsymbol{\Sigma}}_{\mathrm{res}}$ make explicit use of the response and $\widehat{\boldsymbol{\Sigma}}$ does not, the response still enters when using $\widehat{\boldsymbol{\Sigma}}$ because we regress $\boldsymbol{\Sigma}^{-1/2}\mathbf{X}_y$ on $\mathbf{f}_y$ to obtain the principal fitted components. Any of these five form can be used in practice since they each give the same estimated subspace, but we tend to use $\mathcal{S}_d(\widehat{\boldsymbol{\Sigma}}_{\mathrm{res}}, \widehat{\boldsymbol{\Sigma}}_{\mathrm{fit}})$ for no compelling reason.

### 3.2 Robustness

In this section we consider the robustness of $\mathcal{S}_d(\widehat{\boldsymbol{\Sigma}}, \widehat{\boldsymbol{\Sigma}}_{\mathrm{fit}})$ as an estimator $\boldsymbol{\Delta}^{-1}\mathcal{S}_{\boldsymbol{\Gamma}}$ under nonnormality of the errors and misspecification of the model for $\boldsymbol{\nu}_y$. Specifically, we still assume that model (1) holds, but now with possibly nonnormal errors that are independent of $Y$ and have finite moments. The fitted model has mean function as given in (2), but we no longer assume that $\boldsymbol{\nu}_y = \boldsymbol{\beta}\{\mathbf{f}_y - \mathrm{E}(\mathbf{f}_Y)\}$. This then allows for misspecification of $\mathbf{f}_y$. Let $\boldsymbol{\rho}$ be the $d \times r$ matrix of correlations between the elements of $\boldsymbol{\nu}_Y$ and $\mathbf{f}_Y$. Then, with this understanding:

THEOREM 3.5. *$\mathcal{S}_d(\widehat{\boldsymbol{\Sigma}}, \widehat{\boldsymbol{\Sigma}}_{\mathrm{fit}})$ is a $\sqrt{n}$ consistent estimator of $\boldsymbol{\Delta}^{-1}\mathcal{S}_{\boldsymbol{\Gamma}}$ if and only if $\boldsymbol{\rho}$ has rank $d$.*

This result indicates that we may still expect $\mathcal{S}_d(\widehat{\boldsymbol{\Sigma}}, \widehat{\boldsymbol{\Sigma}}_{\mathrm{fit}})$ to be a reasonable estimator when $\mathbf{f}_y$ is misspecified, provided that it is sufficiently correlated with $\boldsymbol{\nu}_y$. It also places the present methodology on an equal footing with other $\sqrt{n}$ consistent methods that do not explicitly assume normality at the outset.

While $\sqrt{n}$ consistency does not necessarily guarantee good performance in practice, our experiences with simulations suggest that it is not difficult to choose an adequate $\mathbf{f}_y$. To illustrate we generated $n = 200$ observations from model (1) with $d = 1$, $Y \sim U(0, 4)$, $\boldsymbol{\nu}_y = \exp(y)$, $p = 20$, $\boldsymbol{\Gamma} = (1, \ldots, 1)^T/\sqrt{20}$ and $\boldsymbol{\Delta} = \mathbf{I}_p$. This choice for $\boldsymbol{\Delta}$ involves no loss of generality because of the invariance property in Corollary 3.3. Each data set was fitted with $d = 1$, $\mathbf{f}_y = (y, y^2, \ldots, y^k)^T$ for $k = 1, \ldots, 6$ and with $\mathbf{f}_y = \exp(y)$. At the suggestion of a referee, we also included the lasso regression of $Y$ on $\mathbf{X}$ (Tibshirani, 1996) constructed by using the R library "relaxo" (Meinshausen, 2006), which selects the tuning parameter by cross validation. For each choice of $\mathbf{f}_y$ we computed with $d = 1$ the angle between $\mathcal{S}_d(\widehat{\boldsymbol{\Sigma}}, \widehat{\boldsymbol{\Sigma}}_{\mathrm{fit}})$ and $\boldsymbol{\Delta}^{-1}\mathcal{S}_{\boldsymbol{\Gamma}}$. We also computed the angle between the lasso coefficient vector and $\boldsymbol{\Delta}^{-1}\mathcal{S}_{\boldsymbol{\Gamma}}$. Figure 1 shows boxplots of the angles for 100 replications. For reference, the expected angle between $\boldsymbol{\Delta}^{-1}\mathcal{S}_{\boldsymbol{\Gamma}}$ and a randomly chosen vector in $\mathbb{R}^{20}$ is about 80 degrees. The quadratic $\mathbf{f}_y$ shows considerable improvement over the linear case, and the results for the 3rd through 6th degree are indistinguishable from those for the model used to generate the data.

The lasso performance was better than PFC with $\mathbf{f}_y = y$, but not otherwise. However, the performance of the lasso may improve in sparse regressions with relatively few active predictors. To address this possibility, we repeated the simulations of Figure 1 after setting elements of $\boldsymbol{\Gamma}$ to zero and renormalizing to length 1. The performance of all PFC estimators was essentially the same as those shown in Figure 1. The lasso did improve, but was still noticeably less accurate than PFC with $\mathbf{f}_y = (y, y^2, \ldots, y^k)^T$ and $k \geq 2$. For example, with only five active predictors, the median lasso angle was about 41 degrees. Section 7.2 contains additional discussion of the lasso.

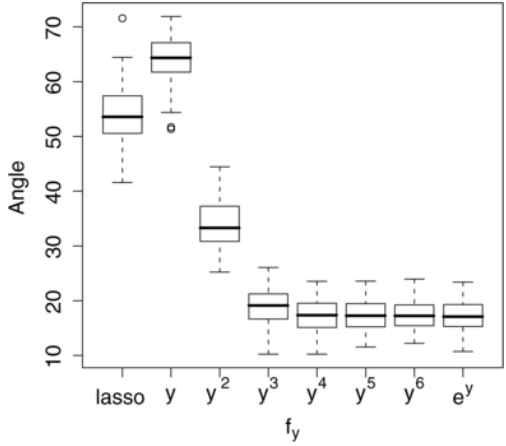

Fig. 1. *Boxplots of the angle between each of eight estimators and* $\Delta^{-1}\mathcal{S}_\Gamma$. *The first boxplot is for the lasso. Boxplots 2–8 are for the PFC estimators* $\mathcal{S}_d(\widehat{\Sigma}, \widehat{\Sigma}_{\mathrm{fit}})$ *under various choices for* $\mathbf{f}_y$: *boxplots 2–7 are labeled according to the last term in* $\mathbf{f}_y = (y, y^2, \ldots, y^k)^T$, $k = 1, \ldots, 6$. *The last boxplot is for* $\mathbf{f}_y = \exp(y)$.

## 4. RELATIONSHIPS WITH OTHER METHODS

### 4.1 Sliced Inverse Regression

Cook (2007) proved that when $Y$ is categorical and $\mathbf{f}_y$ is an indicator vector for the $Y$ categories, the SIR estimator (Li, 1991) of the central subspace is $\mathcal{S}_d(\widehat{\Sigma}, \widehat{\Sigma}_{\mathrm{fit}})$. Theorem 2.1 and Corollary 3.4 imply that, under model (2) with $Y$ categorical, the SIR estimator is the MLE of $\Delta^{-1}\mathcal{S}_\Gamma$. This and Theorem 3.5 help explain many of the empirical findings on the operating characteristics of SIR. If $Y$ is categorical and model (2) is accurate, SIR will inherit optimality properties from general likelihood theory. If $Y$ is continuous and slicing is used then SIR will provide a $\sqrt{n}$ consistent estimator when $\boldsymbol{\rho}$, the matrix of correlations between the elements of $\boldsymbol{\nu}_Y$ and the vector $\mathbf{f}_Y$ of slice indicators, has rank $d$. However, in practice SIRs step functions may provide only a rough approximation to $\mathrm{E}(\mathbf{X}_y)$ and, as a consequence, can leave useful intra slice information behind. While this information might be recovered by intra slice fitting (Cook and Ni, 2006), we expect that PFC modeling is not as prone to such information loss.

### 4.2 Partial and Ordinary Least Squares

To develop connections between PFC, partial least squares (PLS) and ordinary least squares (OLS), consider regressing $Y$ on $\mathbf{X}$ in two steps, assuming that $d = 1$. First we reduce $\mathbf{X}$ linearly to $\mathbf{G}^T\mathbf{X}$ using some methodology that produces $\mathbf{G} \in \mathbb{R}^{p \times q}$, $q \leq p$. The second step consists of using OLS to fit the regression of $Y$ on $\mathbf{G}^T\mathbf{X}$. Letting $\widehat{\boldsymbol{\beta}}_\mathbf{G}$ denote the resulting vector of coefficient for $\mathbf{X}$, we have

$$\widehat{\boldsymbol{\beta}}_\mathbf{G} = \mathbf{P}_{\mathbf{G}(\widehat{\Sigma})}\widehat{\boldsymbol{\beta}}_{\mathrm{ols}} = \mathbf{G}(\mathbf{G}^T\widehat{\Sigma}\mathbf{G})^{-1}\mathbf{G}^T\mathbb{X}^T\mathbb{Y}/n,$$

where $\mathbb{Y}$ is the $n \times 1$ vector of centered responses and $\widehat{\boldsymbol{\beta}}_{\mathrm{ols}}$ is the vector of coefficients from the OLS fit of $Y$ on $\mathbf{X}$. This estimator, which is the projection of $\widehat{\boldsymbol{\beta}}_{\mathrm{ols}}$ onto $\mathrm{span}(\mathbf{G})$ in the $\widehat{\Sigma}$ inner product, does not require computation of $\widehat{\Sigma}^{-1}$ if $q < p$ and thus may be useful when $n < p$, depending on the size of $q$.

Clearly, if $\mathbf{G} = \mathbf{I}_p$ then $\widehat{\boldsymbol{\beta}}_\mathbf{G} = \widehat{\boldsymbol{\beta}}_{\mathrm{ols}}$. Consider next constructing $\mathbf{G}$ from PFC using $\mathbf{f}_y = y$. In that case $\mathrm{span}(\mathbf{G}) = \mathcal{S}_1(\widehat{\Sigma}, \widehat{\Sigma}_{\mathrm{fit}}) = \mathrm{span}(\widehat{\boldsymbol{\beta}}_{\mathrm{ols}})$, and consequently using PFC with $\mathbf{f}_y = y$ to construct $\mathbf{G}$ produces the OLS estimator. The simulations shown in the boxplot of Figure 1 with $\mathbf{f}_y = y$ then correspond to OLS.

Let $\mathbf{C} = \mathrm{cov}(\mathbf{X}, Y)$ and $\widehat{\mathbf{C}} = \mathbb{X}^T\mathbb{Y}/n$. Setting $\mathbf{G} = (\widehat{\mathbf{C}}, \widehat{\Sigma}\widehat{\mathbf{C}}, \ldots, \widehat{\Sigma}^{q-1}\widehat{\mathbf{C}})$ yields the PLS estimator with $q$ factors (Helland, 1990). PLS works best when $\mathbf{C}$ can be expressed as a linear combination of $q$ eigenvectors of $\Sigma$ with unique eigenvalues (Helland and Almøy, 1994), and then $\mathrm{span}(\mathbf{G})$ is an estimator of the span of those eigenvectors. From these results it can be seen that using the isotonic PFC subspace $\mathbf{G} = \mathcal{S}_1(\mathbf{I}_p, \widehat{\Sigma}_{\mathrm{fit}})$ with $\mathbf{f}_y = y$ produces a $\widehat{\boldsymbol{\beta}}_\mathbf{G}$ that is equal to the PLS estimator with $q = 1$. Connections between PLS and PFC are less clear when $q > 1$.

### 4.3 Seeded Reductions

As mentioned in the Introduction, there has been recent interest in principal components as a reductive method for regressions in which $n < p$. Isotonic PFC applies directly when $n < p$, as do the methods with a structured $\Delta$ discussed in Section 8. The PFC estimator with unstructured $\Delta > 0$ is not directly applicable to $n < p$ regressions, but it does provide a seed for recently methodology that is applicable. Cook, Li and Chiaromonte (2007) developed methodology for estimating the central subspace $\mathcal{S}_{Y|\mathbf{X}}$ in $n < p$ regressions when there is a population seed matrix $\boldsymbol{\phi} \in \mathbb{R}^{p \times d}$ such that $\mathrm{span}(\boldsymbol{\phi}) = \Sigma\mathcal{S}_{Y|\mathbf{X}}$. It follows from Corollary 3.4 that, in the context of this article, $\mathcal{S}_{Y|\mathbf{X}} = \boldsymbol{\Delta}^{-1}\mathcal{S}_{\boldsymbol{\Gamma}} = \boldsymbol{\Sigma}^{-1}\mathrm{span}(\boldsymbol{\Sigma}_{\mathrm{fit}})$, where $\boldsymbol{\Sigma}_{\mathrm{fit}}$ is the population version of $\widehat{\boldsymbol{\Sigma}}_{\mathrm{fit}}$. Let the columns of $\boldsymbol{\phi}_{\mathrm{fit}}$ be the eigenvectors of $\boldsymbol{\Sigma}_{\mathrm{fit}}$ corresponding to its $d$ largest eigenvalues. Then $\mathrm{span}(\boldsymbol{\phi}_{\mathrm{fit}}) = \boldsymbol{\Sigma}\mathcal{S}_{Y|\mathbf{X}}$ and $\boldsymbol{\phi}_{\mathrm{fit}}$ qualifies as a population seed matrix. The sample version of $\boldsymbol{\phi}_{\mathrm{fit}}$ is constructed in the same way using $\widehat{\boldsymbol{\Sigma}}_{\mathrm{fit}}$. At this point the methodology of Cook, Li and Chiaromonte (2007) applies directly.

## 5. CHOICE OF $d$

The dimension $d$ of the central subspace was so far assumed to be specified. There are at least two ways to choose $d$ in practice. The first is based on using likelihood ratio statistics $\Lambda_w = 2(L_{\min(p,r)} - L_w) = -n\sum_{i=d+1}^{\min(p,r)}\log(1 - r_i^2)$ to test the null hypothesis $d = w$ against the general alternative $d > w$. Under the null hypothesis $d = w$, $\Lambda_w$ has an asymptotic chi-square distribution with $(r - w)(p - w)$ degrees of freedom. This statistic is the same as the usual likelihood ratio statistic for testing if the last $\min(p,r) - d$ canonical correlations are 0 in two sets of jointly normal random variables (see, e.g., Muirhead, 1982, page 567). In the present application the conditional random vector $\mathbf{X}_y$ is normal, but marginally $\mathbf{X}$ and $\mathbf{f}_Y$ will typically be nonnormal. The likelihood ratio test (LRT) is used sequentially, starting with $w = 0$ and estimating $d$ as the first hypothesized value that is not rejected.

The second approach is to use an information criterion like AIC or BIC. BIC is consistent for $d$ while AIC is minimax-rate optimal (Burnham and Anderson, 2002). For $w \in \{0, \ldots, \min(r,p)\}$, the dimension is selected that minimizes the information criterion $IC(w) = -2L_w + h(n)g(w)$, where $L_w$ was defined in (5), $g(w)$ is the number of parameters to be estimated as a function of $w$, in our case, $p(p+3)/2 + rw + w(p - w)$ and $h(n)$ is equal to $\log n$ for BIC and 2 for AIC. These versions are simple adaptations of the commonly occurring asymptotic forms for other models.

Since the choice of $d$ is essential to the proposed methodology, we next discuss selected results from a simulation study to demonstrate that reasonable inference on $d$ is possible. We first generated $Y \sim N(0, \sigma_y^2)$, and then with $d = 2$ generated $\mathbf{X}_y = \boldsymbol{\Gamma}\boldsymbol{\beta}\mathbf{f}_y + \boldsymbol{\varepsilon}$, where $\boldsymbol{\varepsilon} \sim N(0, \boldsymbol{\Delta})$, $\boldsymbol{\beta} = \mathbf{I}_2$, $\mathbf{f}_y = (y, |y|)^T$ and $\boldsymbol{\Gamma} = (\boldsymbol{\Gamma}_1, \boldsymbol{\Gamma}_2) \in \mathbb{R}^{p \times 2}$, with $\boldsymbol{\Gamma}_1 = (1, 1, -1, -1, 0, \ldots, 0)^T/\sqrt{4}$ and $\boldsymbol{\Gamma}_2 = (1, 0, 1, 0, 1, 0, \ldots, 0)^T/\sqrt{3}$. For each $p$, $\boldsymbol{\Delta}$ was generated once as $\boldsymbol{\Delta} = \mathbf{A}^T\mathbf{A}$, where $\mathbf{A}$ is a $p \times p$ matrix of independent standard normal random variables. Let $F(2)$, $F(2,3)$ and $F(2,3,4)$ denote the fractions of simulation runs in which $d$ was estimated to be one of the integer arguments.

Using $\mathbf{f}_y = (y, |y|, y^3)^T$ when fitting with model (2), Figures 2a–2d give the fraction $F(2)$ of runs in which the indicated procedure selected $d = 2$ versus $n$ for $p = 5$, four values of $\sigma_y$ and the three methods under consideration. The number of repetitions was 500. Here and in other figures the variation in the results for adjacent sample sizes reflects simulation error in addition to systematic trends. The relative performance of the methods in Figure 2a depends on the sample $n$ and signal $\sigma_y$ size, and all three method improve as $n$ and $\sigma_y$ increase.

In Figure 3 $\sigma_y = 2$ and $n = 200$. For Figures 3a and 3c, model (2) was fitted with $\mathbf{f}_y = (y, |y|, y^3)^T$, while for the other two figures fitting was done with $\mathbf{f}_y = (y, |y|, y^3, \ldots, y^{10})^T$. Figures 3a and 3b show, as expected, that the chance of choosing the correct value of $d$ decreases with $p$ for all procedures. Figures 3c and 3d show that, with increasing $p$, LRT and AIC slightly overestimate $d$, while BIC underestimates $d$. In the case of AIC, we estimated nearly a 100 percent chance that the estimated $d$ is 2, 3 or 4 with 80 predictors, $r = 10$ and 200 observations. A little overestimation seems tolerable in the context of this article, since then $\widehat{R}$ will still estimate a sufficient reduction and we can always pursue further refinement based on the subsequent forward regressions. With underestimation $\widehat{R}$ no longer estimates a sufficient reduction. While we believe this to be a useful practical result, it is possible that the development of Bartlett-type corrections will reduce the tendency to overestimate. Based on these and other simulations we judged AIC to be the best overall method for selecting $d$, although in the right situation either of the other methods may perform better. For instance, comparing the results in Figures 3a with the results in Figure 2c, we can see that for $n = 200$ and $p = 5$, the performance of BIC was better than AIC. Nevertheless that is reversed after $p \sim 10$ where AIC consistently gave better results.

## 6. TESTING PREDICTORS

In this section we develop tests for hypotheses of the form $Y \perp\!\!\!\perp \mathbf{X}_2 | \mathbf{X}_1$ where the predictor vector is partitioned as $\mathbf{X} = (\mathbf{X}_1^T, \mathbf{X}_2^T)^T$ with $\mathbf{X}_1 \in \mathbb{R}^{p_1}$ and $\mathbf{X}_2 \in \mathbb{R}^{p_2}$, $p = p_1 + p_2$. Under this hypothesis, $\mathbf{X}_2$ furnishes no information about the response once $\mathbf{X}_1$ is known. The following lemma facilitates the development of a likelihood ratio test statistic under model (2). In preparation, partition $\boldsymbol{\Gamma} = (\boldsymbol{\Gamma}_1^T, \boldsymbol{\Gamma}_2^T)^T$, $\boldsymbol{\Sigma} = (\boldsymbol{\Sigma}_{ij})$, $\boldsymbol{\Delta} = (\boldsymbol{\Delta}_{ij})$, $\widehat{\boldsymbol{\Sigma}}_{\text{res}} = (\widehat{\boldsymbol{\Sigma}}_{ij,\text{res}})$, $\boldsymbol{\Sigma} = (\boldsymbol{\Sigma}_{ij})$, and $\boldsymbol{\Delta}^{-1} = (\boldsymbol{\Delta}^{ij})$, $i = 1, 2$, $j = 1, 2$, to conform to the partitioning of $\mathbf{X}$. Let $\boldsymbol{\Delta}^{-ii} = (\boldsymbol{\Delta}^{ii})^{-1}$. For a square partitioned matrix $\mathbf{A} = (\mathbf{A}_{ij}), i, j = 1, 2$, let $\mathbf{A}_{ii \cdot j} = \mathbf{A}_{ii} - \mathbf{A}_{ij}\mathbf{A}_{jj}^{-1}\mathbf{A}_{ji}$.

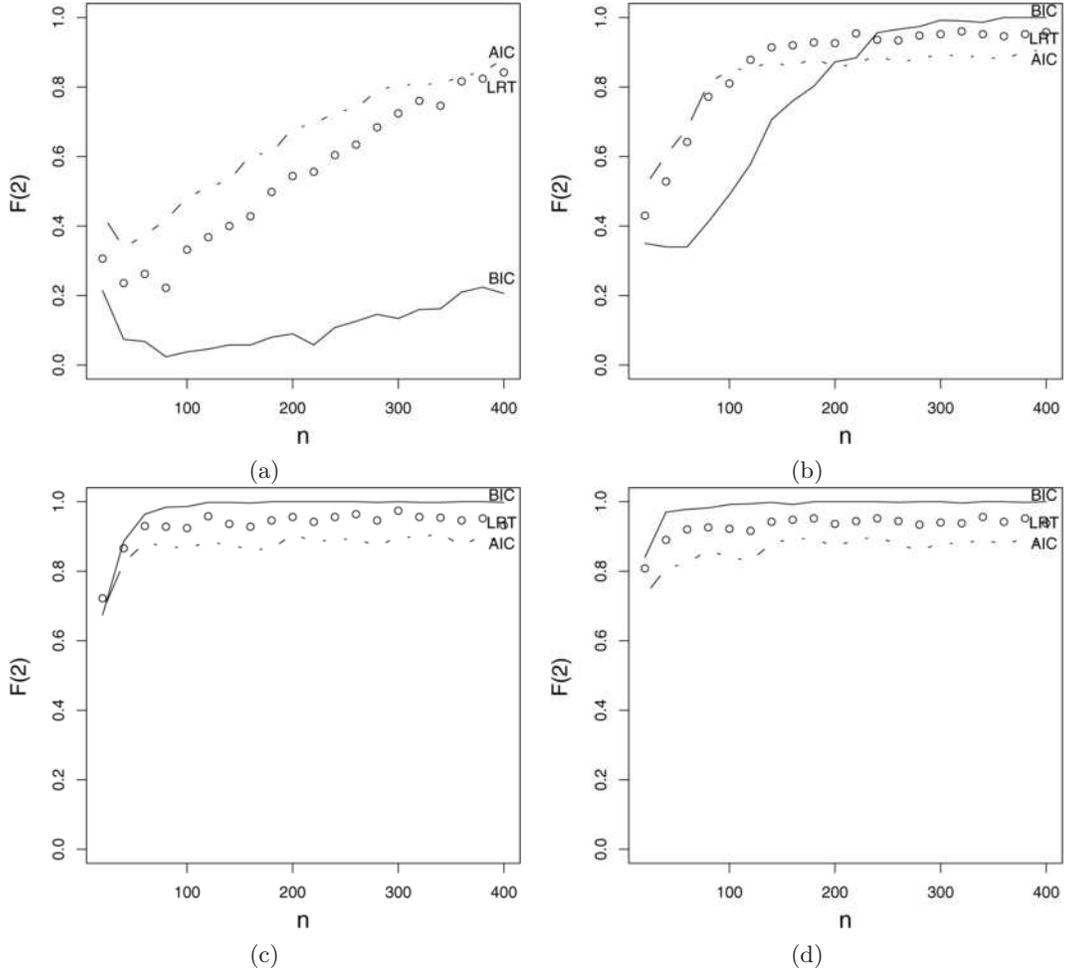

Fig. 2. *Inference about d: Fraction $F(2)$ of replications in which $d = 2$ was chosen by the LRT, AIC and BIC versus the sample size $n$ for four values of $\sigma_y$.* (a) $\sigma_y = 0.5$, (b) $\sigma_y = 1$, (c) $\sigma_y = 2$, (d) $\sigma_y = 5$.

LEMMA 6.1. *Assume model* (2). *Then* $Y \perp\!\!\!\perp \mathbf{X}_2 | \mathbf{X}_1$ *if and only if* $\boldsymbol{\Gamma}_2 = -\boldsymbol{\Delta}^{-22}\boldsymbol{\Delta}^{21}\boldsymbol{\Gamma}_1$.

The log likelihood for the alternative of dependence is as given in Theorem 3.1. The following theorem gives the maximum likelihood estimators under the hypothesis $Y \perp\!\!\!\perp \mathbf{X}_2 | \mathbf{X}_1$.

THEOREM 6.2. *Assume that* $\boldsymbol{\Gamma}_2 = -\boldsymbol{\Delta}^{-22}\boldsymbol{\Delta}^{21}\boldsymbol{\Gamma}_1$ *and that* $d \leq \tau_1 = \min(r, p_1)$. *Then, the MLE of* $\boldsymbol{\Delta}$ *is given in blocks by* $\widehat{\boldsymbol{\Delta}}_{11} = \widehat{\boldsymbol{\Sigma}}_{11,\text{res}}^{1/2} \widehat{\mathbf{V}}(\mathbf{I}_{p_1} + \widehat{\mathbf{K}})\widehat{\mathbf{V}}^T\widehat{\boldsymbol{\Sigma}}_{11,\text{res}}^{1/2}$,

*with* $\widehat{\mathbf{K}} = \text{diag}(0, \ldots, 0, \widehat{\lambda}_{d+1}, \ldots, \widehat{\lambda}_{p_1})$ *and* $\widehat{\mathbf{V}}$ *and* $\widehat{\lambda}_1$, $\ldots, \widehat{\lambda}_{p_1}$ *the ordered eigenvectors and eigenvalues of* $\widehat{\boldsymbol{\Sigma}}_{11,\text{res}}^{-1/2}\widehat{\boldsymbol{\Sigma}}_{11,\text{fit}}\widehat{\boldsymbol{\Sigma}}_{11,\text{res}}^{-1/2}$; $\widehat{\boldsymbol{\Delta}}_{12} = \widehat{\boldsymbol{\Delta}}_{11}\widehat{\boldsymbol{\Sigma}}_{11}^{-1}\widehat{\boldsymbol{\Sigma}}_{12}$ *and* $\widehat{\boldsymbol{\Delta}}_{22} = \widehat{\boldsymbol{\Sigma}}_{22.1} + \widehat{\boldsymbol{\Sigma}}_{21}\widehat{\boldsymbol{\Sigma}}_{11}^{-1}\widehat{\boldsymbol{\Delta}}_{11}\widehat{\boldsymbol{\Sigma}}_{11}^{-1}\widehat{\boldsymbol{\Sigma}}_{12}$. *The MLE of the central subspace is* $\text{span}\{(\widehat{\boldsymbol{\Sigma}}_{11,\text{res}}^{-1/2}\widehat{\mathbf{G}}_1, 0)^T\}$, *with* $\widehat{\mathbf{G}}_1$ *the first* $d$ *eigenvectors of* $\widehat{\boldsymbol{\Sigma}}_{11,\text{res}}^{-1/2}\widehat{\boldsymbol{\Sigma}}_{11,\text{fit}}\widehat{\boldsymbol{\Sigma}}_{11,\text{res}}^{-1/2}$. *The maximum value of the log likelihood is*

$$
\begin{aligned}
L_d^1 = &-\frac{np}{2}\log(2\pi) - \frac{np}{2} - \frac{n}{2}\log|\widehat{\boldsymbol{\Sigma}}_{11,\text{res}}| \\
&- \frac{n}{2}\log|\widehat{\boldsymbol{\Sigma}}_{22.1}| - \frac{n}{2}\sum_{i=d+1}^{\tau_1}\log(1 + \widehat{\lambda}_i).
\end{aligned}
\tag{6}
$$

Under the hypothesis $Y \perp\!\!\!\perp \mathbf{X}_2 | \mathbf{X}_1$ the likelihood ratio statistic $\Theta_d = 2(L_d - L_d^1)$ has an asymptotic chi-squared distribution with $dp_2$ degrees of freedom. By following Corollary 3.2 this test statistic

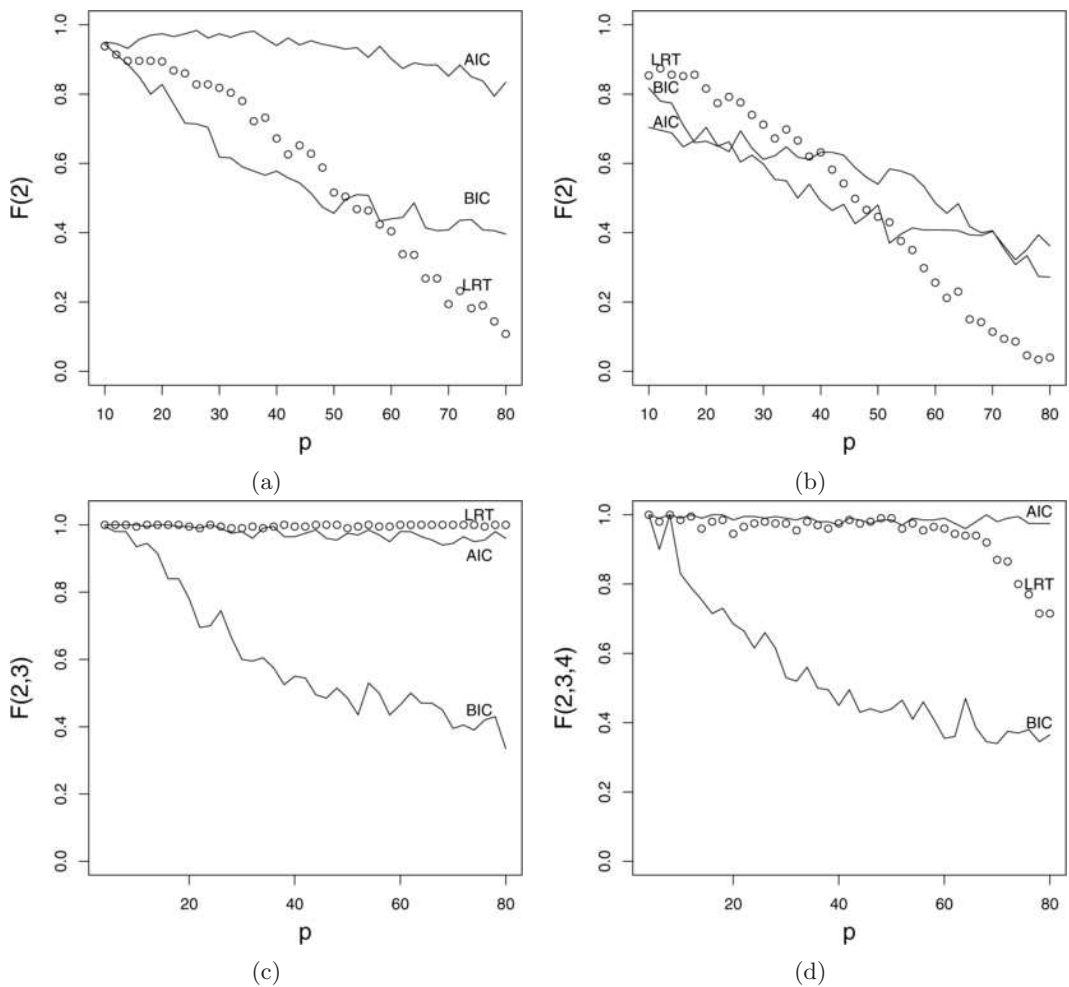

FIG. 3. *Inference about $d$ with varying $p$ and two versions of $\mathbf{f}_y$ used in fitting for LRT, AIC and BIC.* (a) $r = 3$, (b) $r = 10$, (c) $r = 3$, (d) $r = 10$.

can be expressed also as

$$
\begin{aligned}
\Theta_d = {}& n \log |\hat{\boldsymbol{\Sigma}}_{22.1}| - n \log |\hat{\boldsymbol{\Sigma}}_{22.1,\mathrm{res}}| \\
& + n \sum_{i=d+1}^{\min(p,r)} \log(1 - r_i^2) \\
& - n \sum_{i=d+1}^{\min(p_1,r)} \log(1 - t_i^2),
\end{aligned}
$$

where $t_1, \ldots, t_{\min(p_1,r)}$ are the sample canonical correlations between $\mathbf{X}_1$ and $\mathbf{f}_y$.

By using a working dimension $w = \min(r, p_1)$ when constructing the likelihood ratio statistic, we can test predictors without first inferring about $d$, in the same way that we set $w = \min(r, p)$ when testing hypotheses about the structure of $\boldsymbol{\Delta}$. In that case, the final term of $\Theta_d$ does not appear.

To study the proposed tests of $Y \perp\!\!\!\perp \mathbf{X}_2 | \mathbf{X}_1$, we generated data from the model $\mathbf{X}_y = \boldsymbol{\Gamma} y + \boldsymbol{\varepsilon}$, where $\boldsymbol{\varepsilon} \sim N(0, \boldsymbol{\Delta})$ and $Y \sim N(0, \sigma_y^2)$; $\boldsymbol{\Delta}$ was generated as in Section 5, and $\boldsymbol{\Gamma} = c(\boldsymbol{\Gamma}_1^T, \boldsymbol{\Gamma}_2^T)^T \in \mathbb{R}^{10}$ with $\boldsymbol{\Gamma}_1 = (1, \ldots, 1)^T \in \mathbb{R}^7$, $\boldsymbol{\Gamma}_2 = -\boldsymbol{\Delta}^{-22} \boldsymbol{\Delta}^{21} \boldsymbol{\Gamma}_1$, and $c$ is a normalizing constant. Predictor testing is best done after choice of $d$, so the fitted model was (2) with $d = r = 1$ and $\mathbf{f}_y = y$. Partition $\mathbf{X} = (\mathbf{X}_1^T, \mathbf{X}_2^T)^T$ with $\dim(\mathbf{X}_1) = 7$. Our general conclusion from this and other simulations is that the actual and nominal levels of the test are usefully close, except when the sample size $n$ or signal size $\sigma_y$ is quite small. For instance, the estimated levels of nominal 5 percent tests based on 500 runs were 0.18, 0.08, 0.06 and 0.05 for sample sizes 20, 40, 100 and 120. The test tends to reject too frequently for weak signals or small samples. We see that there is again a tendency for likelihood methods to overestimate, in this case the active predictors. As with inference on $d$ we do not judge this to be a serious issue in the context of this article.

## 7. ILLUSTRATIONS

We use two small examples in this section to illustrate the proposed methodology. The first is the most thorough, while the second focuses on selected aspects of the methodology.

### 7.1 Wheat Protein

We use Fearn's ([1983](#)) wheat protein data for this illustration. The response is the protein content of a sample of ground wheat, and the predictors are $-\log(\text{reflectance})$ of NIR radiation at $p = 6$ wavelengths. We chose this example because the predictors are highly correlated, with pairwise sample correlations ranging between 0.9118 and 0.9991. Principal components are often considered in such regressions to mitigate the variance inflation that can be caused by collinearity.

Plots of each predictor versus $Y$ (not shown) suggest that $\mathrm{E}(\mathbf{X}|Y = y)$ might be modeled adequately as a quadratic function of $y$ and thus $\mathbf{f}_y = (y, y^2)^T$, but for this illustration we decided to allow more flexibility and so set $\mathbf{f}_y = (y, y^2, y^3)^T$. Fitting model ([2](#)) with this cubic $\mathbf{f}_y$ resulted in AIC, BIC and LRT all choosing $d = 1$, suggesting that only one linear combination of the predictors $\widehat{R}_1$ is sufficient. The plot in Figure [4](#)a of $Y$ versus $\widehat{R}_1$ shows a strong linear relation. In contrast, there is no relationship evident in the plot of $Y$ versus the first principal component shown in Figure [4](#)b. The first four principal components are needed to linearly reconstruct $\widehat{R}_1$.

Application of the likelihood ratio test $\Theta_d$ of Section [6](#) with $d = 1$ to each predictor gave three small $p$-values $(3.8 \times 10^{-11}, 4.6 \times 10^{-12}, 2.5 \times 10^{-5})$ for the third, fourth and sixth wavelengths. The $p$-values for the remaining three wavelengths were all greater than 0.33. Evidently, not all wavelengths are necessary for estimating protein content. Predictor selection could be continued by using standard likelihood-based methods, including backward elimination or an information criterion.

The estimated PFC is $\widehat{R}_1 = \widehat{\boldsymbol{\eta}}^T \mathbf{X}$, where $\widehat{\boldsymbol{\eta}} = (0.11, 0.11, -0.84, 0.50, -0.01, 0.12)^T$ is normalized to have length 1. Although the predictors with the three largest absolute coefficients are same as those found to be significant, such coefficients are not generally useful indicators of the importance of a predictor. As in linear regression, the coefficients depend on the scales of the predictors. Multiplying the predictors by a diagonal matrix $\mathbf{D}^{-1}$ to give scaled predictors $\mathbf{D}^{-1}\mathbf{X}$ results in new coefficients $\mathbf{D}\widehat{\boldsymbol{\eta}}$ because, from Corollary [3.3](#), the reduction itself is invariant under full-rank linear transformations of $\mathbf{X}$: $\widehat{R}_1 = \widehat{\boldsymbol{\eta}}^T \mathbf{X} = \widehat{\boldsymbol{\eta}}^T \mathbf{D}\mathbf{D}^{-1}\mathbf{X}$. If an informal comparison of coefficients is desirable, then it seems necessary to at least standardize the predictor by choosing the diagonal elements of $\mathbf{D}$ to be the square roots of the diagonal elements of $\widehat{\boldsymbol{\Sigma}}$, $\widehat{\boldsymbol{\Sigma}}_{\text{res}}$ or $\widehat{\boldsymbol{\Delta}}$. Use of $\widehat{\boldsymbol{\Sigma}}$ seems least desirable because it is affected by the signal, $\boldsymbol{\Sigma} = \boldsymbol{\Delta} + \boldsymbol{\Gamma}\boldsymbol{\beta}\,\text{var}(\mathbf{f}_Y)\boldsymbol{\beta}^T\boldsymbol{\Gamma}^T$.

We found no clear indications in this example that the errors deviate significantly from normality. However, even if the errors in model ([2](#)) are not normal or $\mathbf{f}_y$ is misspecified, we would still expect reasonable estimates because of Theorem [3.5](#).

We can use these results to gain insights about the types of regressions in which principal components might be effective and about why they apparently fail in this example. Suppose that there are $d$ eigenvectors of $\boldsymbol{\Delta}$ that span $\mathcal{S}_{\boldsymbol{\Gamma}}$ to a good approximation. This happens when $\boldsymbol{\Delta} = \sigma^2 \mathbf{I}_p$, as in the PC model. Then $\boldsymbol{\Delta}^{-1}\mathcal{S}_{\boldsymbol{\Gamma}} \approx \mathcal{S}_{\boldsymbol{\Gamma}}$, $R(\mathbf{X}) \approx \boldsymbol{\Gamma}^T \mathbf{X}$ and there is a $d \times d$ orthogonal matrix $\mathbf{O}$ so that the columns of $\boldsymbol{\Gamma}\mathbf{O}$ are approximately eigenvectors of $\boldsymbol{\Delta}$. It then follows that there are $d$ eigenvectors of $\boldsymbol{\Sigma}$ that approximately span $\mathcal{S}_{\boldsymbol{\Gamma}}$. If the signal represented by $\boldsymbol{\beta}\,\text{var}(\mathbf{f}_Y)\boldsymbol{\beta}^T$ is sufficiently strong then these should be the first $d$ eigenvectors of $\boldsymbol{\Sigma}$ with relatively large eigenvalues. In short, if the signal is sufficiently strong and the eigenvectors of $\boldsymbol{\Delta}$ cooperate, then $\boldsymbol{\Sigma}$ will exhibit collinearity in the direction of $\boldsymbol{\Gamma}$. The reverse implication does not necessarily hold, however. As the present illustration shows, collinearity in $\boldsymbol{\Sigma}$ does not necessarily imply that it has $d$ eigenvectors that span $\mathcal{S}_{\boldsymbol{\Gamma}}$ to a useful approximation. Additionally, the correlations between the errors $\boldsymbol{\varepsilon}$ estimated from $\widehat{\boldsymbol{\Delta}}$ range between 0.911 and 0.9993, so the observed collinearity in $\widehat{\boldsymbol{\Sigma}}$ is coming largely from $\widehat{\boldsymbol{\Delta}}$ and does not reflect a strong signal.

### 7.2 Naphthalene

These data consist of 80 observations from an experiment to study the effects of three process variables in the vapor phase oxidation of naphthalene (Franklin et al., [1956](#)). The response is the percentage conversion of naphthalene to naphthoquinone, and the three predictors are air to naphthalene ratio, contact time and bed temperature. Although there are only three predictors in this regression, dimension reduction may still be useful for visualization, as discussed in the Introduction.

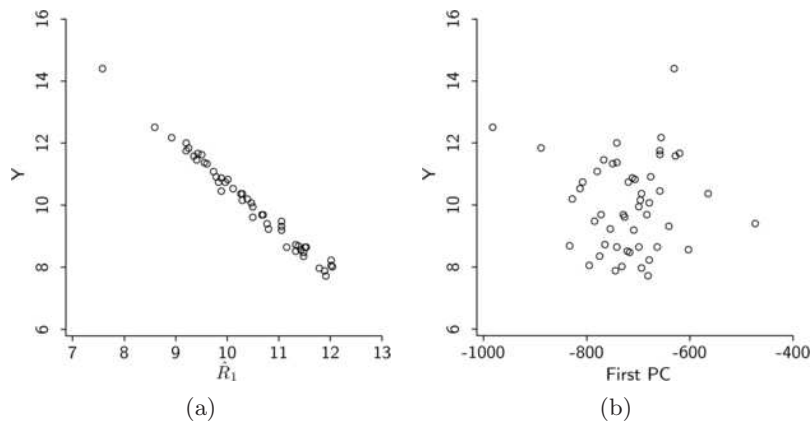

Fig. 4. *Wheat protein data: (a) Response versus the first sufficient component; (b) Response versus the first principal component.*

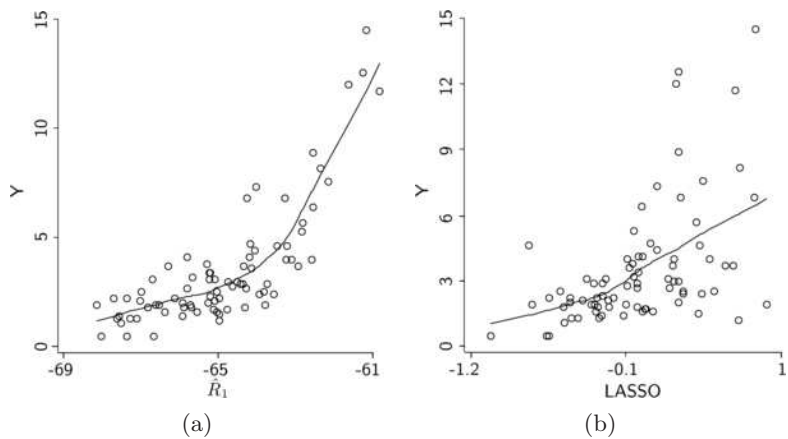

Fig. 5. *Naphthalene data: (a) Response versus the first PFC component; (b) Response versus the lasso fitted values.*

Based on smoothed plots of the predictors versus the response, we used $\mathbf{f}_y = (y, y^2)^T$ to fit the PFC model. The three methods for selecting $d$ discussed in Section 5 all chose $d = 1$. Figure 5a gives a plot of response versus the first principal fitted component $\hat{R}_1$. A plot of the response versus the first principal components failed to show any useful relationship, as in the wheat protein data. We also included in Figure 5b a plot of the response versus lasso fitted values. A plot of the response versus the PLS fitted values with $q = 1$ is quite similar to that shown in Figure 5b.

The lasso and similar penalization methods are designed to fit a single linear combination of the predictors while forcing some of the coefficients to 0, and thereby provide predictor selection along with the fit. If $d = 1$ and the forward linear model is accurate then the lasso should perform well. In the wheat protein data the lasso fitted values are essentially the same as those from PFC shown in Figure 4. If $d = 1$ and the forward linear model is not accurate, then PFC and the lasso can give quite different summaries of the data, as illustrated in Figure 5. Evidently the lasso favors projections of the data that have linear mean functions, and tends to neglect projections with nonlinear mean functions.

The are many examples in the literature where the dimension of the central subspace was inferred to be larger than 1 (see, e.g., Cook, 1998). As presently designed, the lasso cannot respond to such regressions since it fits a single linear combination of the predictors. Similar comments hold for partial least squares and other methods that are constrained by fitting one linear combination of the predictors.

As presently developed, penalization methods like the lasso do not address the issues that drive sufficient dimension reduction. Relatedly, sufficient dimension reduction methods are not designed for automated predictor selection per se. Nevertheless, there is nothing in principle to prohibit using penalization methods within the context of sufficient dimension reduction in an effort to gain the best of both worlds. One might proceed in the context of this article by adding a penalty in $\Delta^{-1}\mathcal{S}_{\Gamma}$ to the partially maximized log likelihood (3).

## 8. STRUCTURED $\Delta$

We would expect the previous methodology to be the most useful in practice. Nevertheless, models between the PFC model with $\Delta = \sigma^2\mathbf{I}_p$ and the PFC model with $\Delta > 0$ may be useful in some applications. In this section we consider models that allow, for example, $\Delta$ to be a diagonal matrix. This will result in a rescaling of the predictors prior to component computation, although that scaling is not the same as the common scaling by marginal standard deviations to produce a correlation matrix. The models discussed here may involve substantially fewer parameters, perhaps resulting in notable efficiency gains when they are reasonable.

Following Anderson (1969) and Rogers and Young (1977), we consider modeling $\Delta$ with a linear structure: $\Delta = \sum_{i=1}^{m}\delta_i\mathbf{G}_i$, where $m \le p(p+1)/2$, $\mathbf{G}_1, \ldots, \mathbf{G}_m$ are known real symmetric $p \times p$ linearly independent matrices and the elements of $\boldsymbol{\delta} = (\delta_1, \ldots, \delta_m)^T$ are functionally independent. We require also that $\Delta^{-1}$ have the same linear structure as $\Delta$: $\Delta^{-1} = \sum_{i=1}^{m} s_i\mathbf{G}_i$. To model a diagonal $\Delta$ we set $\mathbf{G}_i = \mathbf{e}_i\mathbf{e}_i^T$, where $\mathbf{e}_i \in \mathbb{R}^p$ contains a 1 in the $i$th position and zeros elsewhere. This basic structure can be modified straightforwardly to allow for a diagonal $\Delta$ with sets of equal diagonal elements, and for a nondiagonal $\Delta$ with equal off-diagonal entries and equal diagonal entries. In the latter case, there are only two matrices $\mathbf{G}_1 = \mathbf{I}_p$ and $\mathbf{G}_2 = \mathbf{e}\mathbf{e}^T$, where $\mathbf{e} \in \mathbb{R}^p$ has all elements equal to 1.

Estimation of the central subspace $\Delta^{-1}\mathcal{S}_{\Gamma}$ with a constrained $\Delta$ follows that of Section 3 up to Theorem 3.1. The change is that $\Delta$ is now a function of $\boldsymbol{\delta} \in \mathbb{R}^m$. Thus $L_d\{\Delta(\boldsymbol{\delta})\}$ (4) is to be maximized over $\boldsymbol{\delta}$. In contrast to the case with a general $\Delta$, here we were unable to find a closed-form solution to the maximization problem, but any of the standard nonlinear optimization methods should be sufficient to find $\arg\max_{\boldsymbol{\delta}} L\{\Delta(\boldsymbol{\delta})\}$ numerically. We have used an algorithm (Appendix B) to solve $\partial L\{\Delta(\boldsymbol{\delta})\}/\partial\boldsymbol{\delta} = 0$ iteratively. The starting point is the value that maximizes $L_d$ when $r = d$ since then the maximum can be found explicitly. The resulting estimator of the central subspace can be described as $\mathcal{S}_d(\widetilde{\Delta}, \widehat{\Sigma}_{\text{fit}})$, where $\widetilde{\Delta}$ is the MLE of the constrained $\Delta$, but Corollary 3.4 no longer holds.

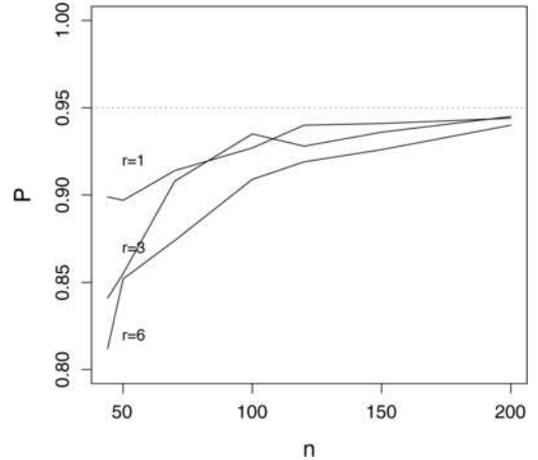

Fig. 6. *Tests of a diagonal* $\Delta$: *The x-axis represents sample size and the y-axis represents the fraction* $P$ *of time the null hypothesis is not rejected.*

A model with constrained $\Delta$ can be tested against (2) by using a likelihood ratio test: under the constrained model $\Omega_d = 2\{L_d - L_d(\widetilde{\Delta})\}$ is distributed asymptotically as a chi-squared random variable with $p(p+1)/2 - m$ degrees of freedom. This test requires that $d$ be specified first, but in practice it may be useful to infer about $\Delta$ prior to inference about $d$. This can be accomplished with some loss of power by overfitting the conditional mean and using the statistic $\Omega_{\min(r,p)}$, which has the same asymptotic null distribution as $\Omega_d$.

To confirm our asymptotic calculations, we generated data from the simulation model $\mathbf{X}_y = \mathbf{\Gamma}y + \boldsymbol{\varepsilon}$, with $Y \sim N(0,1)$, $\mathbf{\Gamma} = (1, \ldots, 1)^T/\sqrt{p} \in \mathbb{R}^p$ and $\boldsymbol{\varepsilon} \sim N(0, \Delta)$, where $\Delta$ is a diagonal matrix with entry $(i,i)$ equal to $10^{i-1}$. For the fitted model we used the working dimension $w = r$, since inference on $\Delta$ will likely be made prior to inference on $d$, and $\mathbf{f}_y = (y, \ldots, y^r)^T$. Testing was done at the 5 percent level and the number of repetitions was 500. Figure 6 gives graphs of the fraction of runs in which the null hypothesis was not rejected versus sample size for various values of $r$ and $p = 6$. These and other simulation results show that the test performs as expected when $n$ is large relative to $p$. As indicated in Figure 6, our simulation results indicate that, with $d$ fixed, the sample size needed to obtain good agreement between the nominal and actual levels of the test increases with $r$.

## 9. DISCUSSION

The methods proposed in this article provide likelihood-based solutions to the two long-standing problems that have hounded principal components, establish a likelihood-based connection between principal fitted components and model-free sufficient dimension reduction and provide insights about the types of regressions in which principal components might be useful. When model (2) is accurate, the methodology will inherit optimality properties from general likelihood theory, while otherwise providing $\sqrt{n}$ consistent estimators under relatively weak conditions. Additionally, there are no restrictions on the nature of the response, which may be continuous, categorical or even multivariate. Perhaps the main limitations are that $\mathrm{var}(\mathbf{X}|Y)$ must be constant or approximately so, and the methods are not designed for discrete or categorical predictors. Investigations into extensions that address these limitations are in progress (Cook and Forzani, 2009).

## APPENDIX A: PROOFS OF THE RESULTS FROM SECTIONS 3 AND 6

PROOF OF THEOREM 2.1. The condition $Y|\mathbf{X} \sim Y|T$ holds if and only if $\mathbf{X}|(T,Y) \sim \mathbf{X}|T$. Thus, thinking of $Y$ as the parameter and $\mathbf{X}$ as the data, $T$ can be regarded as a sufficient statistic for $\mathbf{X}|Y$. The conclusion will follow if we can show that $R$ is a minimal sufficient statistic for $\mathbf{X}|Y$. Note that in this treatment the actual unknown parameters $(\boldsymbol{\mu}, \mathcal{S}_{\Gamma}, \boldsymbol{\beta}, \boldsymbol{\Delta})$ play no essential role.

Let $g(\mathbf{x}|y)$ denote the conditional density of $\mathbf{X}|(Y = y)$. To show that $R$ is a minimal sufficient statistic for $\mathbf{X}|Y$ it is sufficient to consider the log likelihood ratio

$$\log g(\mathbf{z}|y)/g(\mathbf{x}|y)$$
$$= -(1/2)\mathbf{z}^T \boldsymbol{\Delta}^{-1} \mathbf{z} + (1/2)\mathbf{x}^T \boldsymbol{\Delta}^{-1} \mathbf{x}$$
$$+ (\mathbf{z} - \mathbf{x})^T \boldsymbol{\Delta}^{-1} \boldsymbol{\mu}_y.$$

If $\log g(\mathbf{z}|y)/g(\mathbf{x}|y)$ is to be a constant in $y$ then we must have $\log g(\mathbf{z}|y)/g(\mathbf{x}|y) = \mathrm{E}\{\log g(\mathbf{z}|Y)/g(\mathbf{x}|Y)\}$ for all $y$. Equivalently, we must have $(\mathbf{z} - \mathbf{x})^T \boldsymbol{\Delta}^{-1} \cdot (\boldsymbol{\mu}_y - \boldsymbol{\mu}_Y) = 0$. Let $\boldsymbol{\Gamma} \in \mathbb{R}^{p \times d}$ be a basis for span($\boldsymbol{\mu}_y - $

$\boldsymbol{\mu}_Y$). Then the condition can be expressed equivalently as $(\mathbf{z} - \mathbf{x})^T \boldsymbol{\Gamma} \boldsymbol{\beta} \mathbf{f}_y = 0$, and the conclusion follows. □

Let $\mathbb{S}_q^+$ denote the set of $q \times q$ positive definite matrices.

PROOF OF THEOREM 3.1. We use $f$ as a generic function whose definition changes and is given in context. We will make a series of changes of variables to rewrite the problem. Let $\mathbf{U} = \widehat{\boldsymbol{\Sigma}}_{\mathrm{res}}^{1/2} \boldsymbol{\Delta}^{-1} \widehat{\boldsymbol{\Sigma}}_{\mathrm{res}}^{1/2}$ so that maximizing (4) is equivalent to maximizing

$$(7) \quad f(\mathbf{U}) = \log |\mathbf{U}| - \mathrm{tr}(\mathbf{U})$$
$$- \sum_{i=d+1}^{p} \lambda_i (\mathbf{U} \widehat{\boldsymbol{\Sigma}}_{\mathrm{res}}^{-1/2} \widehat{\boldsymbol{\Sigma}}_{\mathrm{fit}} \widehat{\boldsymbol{\Sigma}}_{\mathrm{res}}^{-1/2}).$$

Let $\tau = \min(r, p)$ and use the singular value decomposition to write $\widehat{\boldsymbol{\Sigma}}_{\mathrm{res}}^{-1/2} \widehat{\boldsymbol{\Sigma}}_{\mathrm{fit}} \widehat{\boldsymbol{\Sigma}}_{\mathrm{res}}^{-1/2} = \widehat{\mathbf{V}} \widehat{\boldsymbol{\Lambda}}_\tau \widehat{\mathbf{V}}^T$ where $\widehat{\mathbf{V}} \in \mathbb{R}^{p \times p}$ is an orthogonal matrix and $\widehat{\boldsymbol{\Lambda}}_\tau = \mathrm{diag}(\widehat{\lambda}_1, \ldots, \widehat{\lambda}_\tau, 0, \ldots, 0)$, with $\widehat{\lambda}_1 > \widehat{\lambda}_2 > \cdots > \widehat{\lambda}_\tau > 0$. Calling $\mathbf{H} = \widehat{\mathbf{V}}^T \mathbf{U} \widehat{\mathbf{V}} \in \mathbb{S}_p^+$, (7) becomes

$$(8) \quad f(\mathbf{H}) = \log |\mathbf{H}| - \mathrm{tr}(\mathbf{H}) - \sum_{i=d+1}^{\tau} \lambda_i (\mathbf{H} \widehat{\boldsymbol{\Lambda}}_\tau).$$

We now partition $\mathbf{H}$ as $\mathbf{H} = (\mathbf{H}_{ij})_{i,j=1,2}$, with $\mathbf{H}_{11} \in \mathbb{S}_\tau^+$, $\mathbf{H}_{22} \in \mathbb{S}_{p-\tau}^+$ [for $p = \tau$ we take $\mathbf{H} = \mathbf{H}_{11}$ and go directly to (9)]. Consider the transformation $\mathbb{S}_p^+$ to the space $\mathbb{S}_\tau^+ \times \mathbb{S}_{p-\tau}^+ \times \mathbb{R}^{\tau \times (p-\tau)}$ given by $\mathbf{V}_{11} = \mathbf{H}_{11}$, $\mathbf{V}_{22} = \mathbf{H}_{22} - \mathbf{H}_{12}^T \mathbf{H}_{11}^{-1} \mathbf{H}_{12}$ and $\mathbf{V}_{12} = \mathbf{H}_{11}^{-1} \mathbf{H}_{12}$. This transformation is one to one and onto (Eaton, 1983, Proposition 5.8). As a function of $\mathbf{V}_{11}$, $\mathbf{V}_{22}$ and $\mathbf{V}_{12}$, (8) can be written as

$$\log |\mathbf{V}_{11}||\mathbf{V}_{22}| - \mathrm{tr}(\mathbf{V}_{11}) - \mathrm{tr}(\mathbf{V}_{22})$$
$$- \mathrm{tr}(\mathbf{V}_{12}^T \mathbf{V}_{11} \mathbf{V}_{12}) - \sum_{i=d+1}^{\tau} \lambda_i (\mathbf{V}_{11} \tilde{\boldsymbol{\Lambda}}_\tau),$$

where $\tilde{\boldsymbol{\Lambda}}_\tau = \mathrm{diag}(\widehat{\lambda}_1, \ldots, \widehat{\lambda}_\tau)$, and we have used the fact that the nonzero eigenvalues of $\mathbf{H} \widehat{\boldsymbol{\Lambda}}_\tau$ are the same as those of $\mathbf{H}_{11} \tilde{\boldsymbol{\Lambda}}_\tau$. The term $-\mathrm{tr}(\mathbf{V}_{12}^T \mathbf{V}_{11} \mathbf{V}_{12})$ is the only one that depends on $\mathbf{V}_{12}$. Since $\mathbf{V}_{11}$ is positive definite, $\mathbf{V}_{12}^T \mathbf{V}_{11} \mathbf{V}_{12}$ is positive semidefinite. Thus, the maximum occurs when $\mathbf{V}_{12} = 0$. This implies that $\mathbf{H}_{12} = 0$, $\mathbf{H}_{11} = \mathbf{V}_{11}$, $\mathbf{H}_{22} = \mathbf{V}_{22}$, and we next need to maximize

$$f(\mathbf{H}_{11}, \mathbf{H}_{22})$$
$$= \log |\mathbf{H}_{11}| + \log |\mathbf{H}_{22}|$$
$$- \mathrm{tr}(\mathbf{H}_{11}) - \mathrm{tr}(\mathbf{H}_{22}) - \sum_{i=d+1}^{\tau} \lambda_i (\mathbf{H}_{11} \tilde{\boldsymbol{\Lambda}}_\tau).$$

This function is maximized over $\mathbf{H}_{22}$ at $\mathbf{H}_{22} = \mathbf{I}_{p-\tau}$, then we need to maximize

$$
(9) \qquad \begin{aligned} f(\mathbf{H}_{11}) &= \log|\mathbf{H}_{11}| - \operatorname{tr}(\mathbf{H}_{11}) \\ &\quad - \sum_{i=d+1}^{\tau} \lambda_i(\mathbf{H}_{11}\tilde{\mathbf{\Lambda}}_\tau). \end{aligned}
$$

Letting $\mathbf{Z} = \tilde{\mathbf{\Lambda}}_\tau^{1/2}\mathbf{H}_{11}\tilde{\mathbf{\Lambda}}_\tau^{1/2}$ leads us to maximize $f(\mathbf{Z}) = \log|\mathbf{Z}| - \operatorname{tr}(\mathbf{Z}\tilde{\mathbf{\Lambda}}_\tau^{-1}) - \sum_{i=d+1}^{\tau} \lambda_i(\mathbf{Z})$. Since $\mathbf{Z} \in \mathbf{S}_\tau^+$, there exists an $\mathbf{F} = \operatorname{diag}(f_1,\ldots,f_\tau)$ with $f_i > 0$ in decreasing order and an orthogonal matrix $\mathbf{W}$ in $\mathbb{R}^{\tau\times\tau}$ such that $\mathbf{Z} = \mathbf{W}^T\mathbf{F}\mathbf{W}$. As a function of $\mathbf{W}$ and $\mathbf{F}$, we can rewrite the function $f$ as

$$
\begin{aligned} f(\mathbf{F}, \mathbf{W}) &= \log|\mathbf{F}| - \operatorname{tr}(\mathbf{W}^T\mathbf{F}\mathbf{W}\tilde{\mathbf{\Lambda}}_\tau^{-1}) - \sum_{i=d+1}^{\tau} f_i \\ &= \log|\mathbf{F}| - \operatorname{tr}(\mathbf{F}\mathbf{W}\tilde{\mathbf{\Lambda}}_\tau^{-1}\mathbf{W}^T) - \sum_{i=d+1}^{\tau} f_i. \end{aligned}
$$

Now, using a lemma from Anderson ([1971](#)), Theorem A.4.7, $\min_{\mathbf{W}} \operatorname{tr}(\mathbf{F}\mathbf{W}\tilde{\mathbf{\Lambda}}_\tau^{-1}\mathbf{W}^T) = \sum_{i=1}^{\tau} f_i\hat{\lambda}_i^{-1}$, and if the diagonal element of $\mathbf{F}$ and $\tilde{\mathbf{\Lambda}}_\tau$ are distinct, the minimum occur when $\widehat{\mathbf{W}} = \mathbf{I}_\tau$. Knowing this, we can rewrite the problem one last time, as that of maximizing in $(f_1,\ldots,f_\tau)$, all greater than zero, the function

$$
(10) \qquad \begin{aligned} & f(f_1,\ldots,f_\tau) \\ &= \sum_{i=1}^{\tau} \log f_i - \sum_{i=1}^{\tau} f_i\hat{\lambda}_i^{-1} - \sum_{i=d+1}^{\tau} f_i. \end{aligned}
$$

Clearly the maximum will occur at $f_i = \hat{\lambda}_i$ for $i = 1,\ldots,d$ and for $i = d+1,\ldots,\tau$, $f_i = \hat{\lambda}_i/(\hat{\lambda}_i+1)$. Since $\hat{\lambda}_i$ are positive and decreasing order, $f_i$ are positive and decreasing in order. Since all the $\hat{\lambda}_i$ are different, the $f_i$ are different. Collecting all the results, the value of $\mathbf{\Delta}$ that maximizes ([4](#)) is

$$
\begin{aligned} \widehat{\mathbf{\Delta}} &= \widehat{\mathbf{\Sigma}}_{\text{res}}^{1/2}\widehat{\mathbf{U}}^{-1}\widehat{\mathbf{\Sigma}}_{\text{res}}^{1/2} = \widehat{\mathbf{\Sigma}}_{\text{res}}^{1/2}\widehat{\mathbf{V}}\widehat{\mathbf{H}}^{-1}\widehat{\mathbf{V}}^T\widehat{\mathbf{\Sigma}}_{\text{res}}^{1/2} \\ &= \widehat{\mathbf{\Sigma}}_{\text{res}}^{1/2}\widehat{\mathbf{V}}\begin{pmatrix} \tilde{\mathbf{\Lambda}}_\tau^{1/2}\widehat{\mathbf{Z}}^{-1}\tilde{\mathbf{\Lambda}}_\tau^{1/2} & \mathbf{0}_{\tau\times(p-\tau)} \\ \mathbf{0}_{(p-\tau)\times\tau} & \mathbf{I}_{p-\tau\times(p-\tau)} \end{pmatrix}\widehat{\mathbf{V}}^T\widehat{\mathbf{\Sigma}}_{\text{res}}^{1/2}, \end{aligned}
$$

where $\tilde{\mathbf{\Lambda}}_\tau^{1/2}\widehat{\mathbf{Z}}^{-1}\tilde{\mathbf{\Lambda}}_\tau^{1/2} = \operatorname{diag}(\mathbf{I}_d, \hat{\lambda}_{d+1}+1,\ldots,\hat{\lambda}_\tau+1)$.

Now, to obtained the maximum value we replace $\mathbf{\Delta}$ by $\widehat{\mathbf{\Delta}}$ in ([4](#)),

$$
(11) \qquad \begin{aligned} L_d(\widehat{\mathbf{\Delta}}) &= -\frac{np}{2}\log(2\pi) - \frac{n}{2}\log|\widehat{\mathbf{\Delta}}| \\ &\quad - \frac{n}{2}\operatorname{tr}(\widehat{\mathbf{\Delta}}^{-1}\widehat{\mathbf{\Sigma}}_{\text{res}}) - \frac{n}{2}\sum_{i=d+1}^{\tau}\lambda_i(\widehat{\mathbf{\Delta}}^{-1}\widehat{\mathbf{\Sigma}}_{\text{fit}}). \end{aligned}
$$

Since the trace and the eigenvalues are cyclic operations,

$$
(12) \qquad \begin{aligned} \operatorname{tr}(\widehat{\mathbf{\Delta}}^{-1}\widehat{\mathbf{\Sigma}}_{\text{res}}) &= \operatorname{tr}(\widehat{\mathbf{\Delta}}^{-1/2}\widehat{\mathbf{\Sigma}}_{\text{res}}\widehat{\mathbf{\Delta}}^{-1/2}) \\ &= \operatorname{tr}\{\widehat{\mathbf{V}}(\mathbf{I}_p + \widehat{\mathbf{K}})^{-1}\widehat{\mathbf{V}}^T\} \\ &= d + \sum_{i=d+1}^{\tau} 1/(1+\hat{\lambda}_i) + (p-\tau), \end{aligned}
$$

$$
(13) \qquad \begin{aligned} \sum_{i=d+1}^{\tau}\lambda_i(\widehat{\mathbf{\Delta}}^{-1}\widehat{\mathbf{\Sigma}}_{\text{fit}}) &= \sum_{i=d+1}^{\tau}\lambda_i\{\widehat{\mathbf{V}}(\mathbf{I}+\widehat{\mathbf{K}})^{-1} \\ &\qquad \cdot \widehat{\mathbf{V}}^T\widehat{\mathbf{\Sigma}}_{\text{res}}^{-1/2}\widehat{\mathbf{\Sigma}}_{\text{fit}}\widehat{\mathbf{\Sigma}}_{\text{res}}^{-1/2}\} \\ &= \sum_{i=d+1}^{\tau}\hat{\lambda}_i\{\widehat{\mathbf{V}}(\mathbf{I}+\widehat{\mathbf{K}})^{-1}\widehat{\mathbf{V}}^T\widehat{\mathbf{V}}\widehat{\mathbf{K}}\widehat{\mathbf{V}}^T\} \\ &= \sum_{i=d+1}^{\tau}\hat{\lambda}_i\{(\mathbf{I}+\widehat{\mathbf{K}})^{-1}\widehat{\mathbf{K}}\} \\ &= \sum_{i=d+1}^{\tau}\frac{\hat{\lambda}_i}{1+\hat{\lambda}_i}. \end{aligned}
$$

Since $\widehat{\mathbf{\Sigma}}_{\text{res}} > 0$ we have

$$
(14) \qquad \begin{aligned} \log|\widehat{\mathbf{\Delta}}| &= \log|\widehat{\mathbf{\Sigma}}_{\text{res}}^{1/2}\widehat{\mathbf{V}}(\mathbf{I}+\widehat{\mathbf{K}})\widehat{\mathbf{V}}^T\widehat{\mathbf{\Sigma}}_{\text{res}}^{1/2}| \\ &= \log|\widehat{\mathbf{\Sigma}}_{\text{res}}| + \sum_{i=d+1}^{\tau}\log(1+\hat{\lambda}_i). \end{aligned}
$$

Plugging ([12](#)), ([14](#)) and ([15](#)) into ([12](#)) we obtain ([5](#)). $\square$

PROOF OF COROLLARY 3.2. The eigenvalues $\hat{\lambda}_i$ of $\widehat{\mathbf{\Sigma}}_{\text{res}}^{-1/2}\widehat{\mathbf{\Sigma}}_{\text{fit}}\widehat{\mathbf{\Sigma}}_{\text{res}}^{-1/2}$ are the same as those of $\widehat{\mathbf{\Sigma}}_{\text{res}}^{-1}\widehat{\mathbf{\Sigma}}_{\text{fit}}$. These eigenvalues are related to the eigenvalues $r_i^2$ of $\widehat{\mathbf{\Sigma}}^{-1}\widehat{\mathbf{\Sigma}}_{\text{fit}}$ by $1 + \hat{\lambda}_i = (1-r_i^2)^{-1}$ (Cook, [2007](#), Appendix 7). Now the eigenvalues of $\widehat{\mathbf{\Sigma}}^{-1}\widehat{\mathbf{\Sigma}}_{\text{fit}}$ are the same as those of

$$
\begin{aligned} \widehat{\mathbf{\Sigma}}^{-1/2}\widehat{\mathbf{\Sigma}}_{\text{fit}}\widehat{\mathbf{\Sigma}}^{-1/2} &= \widehat{\mathbf{\Sigma}}^{-1/2}(\mathbb{X}^T\mathbb{F}/n)(\mathbb{F}^T\mathbb{F}/n)^{-1} \\ &\qquad \cdot (\mathbb{F}^T\mathbb{X}/n)\widehat{\mathbf{\Sigma}}^{-1/2}, \end{aligned}
$$

where $\mathbb{X}^T\mathbb{F}/n$ is the $p \times r$ matrix of sample correlations between $\mathbf{X}$ and $\mathbf{f}$ and $\mathbb{F}^T\mathbb{F}/n$ is the sample covariance matrix of $\mathbf{f}$. $\square$

PROOF OF COROLLARY 3.3. Recall from Theorem 3.1 that $\widehat{\mathbf{\Delta}} = \widehat{\mathbf{\Sigma}}_{\text{res}} + \widehat{\mathbf{\Sigma}}_{\text{res}}^{1/2}\widehat{\mathbf{V}}\widehat{\mathbf{K}}\widehat{\mathbf{V}}^T\widehat{\mathbf{\Sigma}}_{\text{res}}^{1/2}$, where $\widehat{\mathbf{V}}$ contains the eigenvectors of $\mathbf{B} = \widehat{\mathbf{\Sigma}}_{\text{res}}^{-1/2}\widehat{\mathbf{\Sigma}}_{\text{fit}}\widehat{\mathbf{\Sigma}}_{\text{res}}^{-1/2}$. The transformation $\mathbf{X} \to \mathbf{A}\mathbf{X}$ transforms $\mathbf{B} \to \mathbf{O}\mathbf{B}\mathbf{O}^T$, where $\mathbf{O} = (\mathbf{A}\widehat{\mathbf{\Sigma}}_{\text{res}}\mathbf{A}^T)^{-1/2}\mathbf{A}\widehat{\mathbf{\Sigma}}_{\text{res}}^{1/2}$ is an orthogonal matrix. Consequently, under the transformation $\widehat{\mathbf{K}}$ is invariant, $\widehat{\mathbf{V}} \to \mathbf{O}\widehat{\mathbf{V}}$ and $\widehat{\boldsymbol{\Delta}} \to \mathbf{A}\widehat{\boldsymbol{\Delta}}\mathbf{A}^T$. The rest of the proof follows similarly. $\quad\square$

To prove Corollary 3.4 we need a lemma.

LEMMA A.1. *Let* $\tilde{\mathbf{V}} = \widehat{\boldsymbol{\Sigma}}_{\mathrm{res}}^{-1/2}\widehat{\mathbf{V}}\mathbf{M}^{1/2}$, *where* $\mathbf{M} = (\mathbf{I}_p + \widehat{\mathbf{K}})^{-1}$, *with* $\widehat{\mathbf{V}}$ *and* $\widehat{\mathbf{K}}$ *as in Theorem 3.1.* *Then* $\widehat{\boldsymbol{\Delta}}^{1/2}\tilde{\mathbf{V}}$ *are the normalized eigenvectors of* $\widehat{\boldsymbol{\Delta}}^{-1/2}\widehat{\boldsymbol{\Sigma}}_{\mathrm{fit}} \times \widehat{\boldsymbol{\Delta}}^{-1/2}$.

PROOF. From Theorem 3.1,

$$\widehat{\boldsymbol{\Delta}} = \widehat{\boldsymbol{\Sigma}}_{\mathrm{res}} + \widehat{\boldsymbol{\Sigma}}_{\mathrm{res}}^{1/2}\widehat{\mathbf{V}}\widehat{\mathbf{K}}\widehat{\mathbf{V}}^T\widehat{\boldsymbol{\Sigma}}_{\mathrm{res}}^{1/2}$$
$$= \widehat{\boldsymbol{\Sigma}}_{\mathrm{res}}^{1/2}\widehat{\mathbf{V}}(\mathbf{I}_p + \widehat{\mathbf{K}})\widehat{\mathbf{V}}^T\widehat{\boldsymbol{\Sigma}}_{\mathrm{res}}^{1/2}.$$

Then, $\widehat{\boldsymbol{\Delta}}^{-1} = \widehat{\boldsymbol{\Sigma}}_{\mathrm{res}}^{-1/2}\widehat{\mathbf{V}}(\mathbf{I}_p + \widehat{\mathbf{K}})^{-1}\widehat{\mathbf{V}}^T\widehat{\boldsymbol{\Sigma}}_{\mathrm{res}}^{-1/2}$ $= \widehat{\boldsymbol{\Sigma}}_{\mathrm{res}}^{-1/2}\widehat{\mathbf{V}}\mathbf{M}\widehat{\mathbf{V}}^T\widehat{\boldsymbol{\Sigma}}_{\mathrm{res}}^{-1/2}$. Using the fact that $\widehat{\mathbf{V}}$ are the eigenvectors of $\widehat{\boldsymbol{\Sigma}}_{\mathrm{res}}^{-1/2}\widehat{\boldsymbol{\Sigma}}_{\mathrm{fit}}\widehat{\boldsymbol{\Sigma}}_{\mathrm{res}}^{-1/2}$ we get

$$\widehat{\boldsymbol{\Delta}}^{-1}\widehat{\boldsymbol{\Sigma}}_{\mathrm{fit}}\tilde{\mathbf{V}} = \widehat{\boldsymbol{\Sigma}}_{\mathrm{res}}^{-1/2}\widehat{\mathbf{V}}\mathbf{M}\widehat{\mathbf{V}}^T\widehat{\boldsymbol{\Sigma}}_{\mathrm{res}}^{-1/2}\widehat{\boldsymbol{\Sigma}}_{\mathrm{fit}}\widehat{\boldsymbol{\Sigma}}_{\mathrm{res}}^{-1/2}\widehat{\mathbf{V}}\mathbf{M}^{1/2}$$
$$= \widehat{\boldsymbol{\Sigma}}_{\mathrm{res}}^{-1/2}\widehat{\mathbf{V}}\mathbf{M}\boldsymbol{\Delta}_\tau\mathbf{M}^{1/2} = \tilde{\mathbf{V}}\mathbf{M}^{1/2}\boldsymbol{\Delta}_\tau\mathbf{M}^{1/2}$$
$$= \tilde{\mathbf{V}}\mathbf{M}\boldsymbol{\Delta}_\tau,$$

where $\mathbf{M}\boldsymbol{\Delta}_\tau = \mathrm{diag}(\widehat{\lambda}_1, \ldots, \widehat{\lambda}_d, \widehat{\lambda}_{d+1}/(\widehat{\lambda}_{d+1} + 1), \ldots, \widehat{\lambda}_\tau/(\widehat{\lambda}_\tau + 1), 0, \ldots, 0)$. Therefore $\widehat{\boldsymbol{\Delta}}^{-1}\widehat{\boldsymbol{\Sigma}}_{\mathrm{fit}}$ has eigenvalues $\widehat{\lambda}_1, \ldots, \widehat{\lambda}_d, \widehat{\lambda}_{d+1}/(\widehat{\lambda}_{d+1} + 1), \ldots, \widehat{\lambda}_\tau/(\widehat{\lambda}_\tau + 1), 0, \ldots, 0$ with eigenvectors $\tilde{\mathbf{V}}$, and $\widehat{\mathbf{V}}^T\widehat{\boldsymbol{\Delta}}\widehat{\mathbf{V}} = \mathbf{I}_p$. $\quad\square$

PROOF OF COROLLARY 3.4. From the development leading to Theorem 3.1, the MLE of $\mathrm{span}(\boldsymbol{\Delta}^{-1}\boldsymbol{\Gamma})$ is $\mathcal{S}_d(\widehat{\boldsymbol{\Delta}}, \widehat{\boldsymbol{\Sigma}}_{\mathrm{fit}})$, which establishes the first form. Now, from Lemma A.1, span of the first $d$ columns of $\widehat{\boldsymbol{\Delta}}^{-1/2}\widehat{\boldsymbol{\Delta}}^{1/2}\tilde{\mathbf{V}} = \tilde{\mathbf{V}}$ is the MLE for $\mathrm{span}(\boldsymbol{\Delta}^{-1}\boldsymbol{\Gamma})$. Since $\tilde{\mathbf{V}} = \widehat{\boldsymbol{\Sigma}}_{\mathrm{res}}^{-1/2}\widehat{\mathbf{V}}\mathbf{M}^{1/2}$ and $\mathbf{M}$ is diagonal full rank with the first $d$ elements equal 1, the span of the first $d$ columns of $\tilde{\mathbf{V}}$ is the same of the first $d$ columns of $\widehat{\boldsymbol{\Sigma}}_{\mathrm{res}}^{-1/2}\widehat{\mathbf{V}}$ where $\widehat{\mathbf{V}}$ are the eigenvectors of $\widehat{\boldsymbol{\Sigma}}_{\mathrm{res}}^{-1/2}\widehat{\boldsymbol{\Sigma}}_{\mathrm{fit}}\widehat{\boldsymbol{\Sigma}}_{\mathrm{res}}^{-1/2}$. This prove the fourth form. The proof of the fifth form can be found in Cook (2007) and it follows from the fact that the eigenvectors of $\widehat{\boldsymbol{\Sigma}}^{-1}\widehat{\boldsymbol{\Sigma}}_{\mathrm{fit}}$ and $\widehat{\boldsymbol{\Sigma}}_{\mathrm{res}}^{-1}\widehat{\boldsymbol{\Sigma}}_{\mathrm{fit}}$ are identically, with corresponding eigenvalues $\widehat{\lambda}_i$ and $\widehat{\lambda}_i/(1 - \widehat{\lambda}_i)$. The corollary follows now from the relation between the eigenvectors of the product of the symmetric matrices $\mathbf{AB}$ and the eigenvectors of $\mathbf{A}^{1/2}\mathbf{BA}^{1/2}$. The second and the third forms follow from the fourth and fifth forms and from the fact that $\widehat{\boldsymbol{\Sigma}} = \widehat{\boldsymbol{\Sigma}}_{\mathrm{res}} + \widehat{\boldsymbol{\Sigma}}_{\mathrm{fit}}$. $\square$

PROOF OF THEOREM 3.5. It is sufficient to consider the limiting behavior of $\widehat{\boldsymbol{\Sigma}}^{-1}\widehat{\boldsymbol{\Sigma}}_{\mathrm{fit}}$, because $\mathcal{S}_d(\widehat{\boldsymbol{\Sigma}}, \widehat{\boldsymbol{\Sigma}}_{\mathrm{fit}}) = \widehat{\boldsymbol{\Sigma}}^{-1/2}\mathrm{span}_d(\widehat{\boldsymbol{\Sigma}}^{-1/2}\widehat{\boldsymbol{\Sigma}}_{\mathrm{fit}}\widehat{\boldsymbol{\Sigma}}^{-1/2}) = \mathrm{span}_d(\widehat{\boldsymbol{\Sigma}}^{-1}\widehat{\boldsymbol{\Sigma}}_{\mathrm{fit}})$, where $\mathrm{span}_d(\mathbf{A})$ indicates the span of the first $d$ eigenvectors of $\mathbf{A}$.

The following statements on large sample behavior follow the general line that Li (1991), Section 5, used in his demonstration of $\sqrt{n}$ consistency for SIR. Since $\widehat{\boldsymbol{\Sigma}}$ is the marginal sample covariance matrix of $\mathbf{X}$, its asymptotic behavior depends only on the true model. It is know that under the stated assumptions $\widehat{\boldsymbol{\Sigma}}$ is a $\sqrt{n}$ consistent estimator of $\boldsymbol{\Sigma} = \boldsymbol{\Gamma}\mathbf{V}\boldsymbol{\Gamma}^T + \boldsymbol{\Delta}$, where $\mathbf{V} = \mathrm{var}(\boldsymbol{\nu}_Y) > 0$. Consequently, $\widehat{\boldsymbol{\Sigma}}^{-1}$ is a $\sqrt{n}$ consistent estimator of $\boldsymbol{\Sigma}^{-1}$. Next, as given in Section 2.1, $\widehat{\boldsymbol{\Sigma}}_{\mathrm{fit}} = (\mathbb{X}^T\mathbb{F}/n)(\mathbb{F}^T\mathbb{F}/n)^{-1}(\mathbb{F}^T\mathbb{X}/n)$ which converges at $\sqrt{n}$ rate to $\boldsymbol{\Sigma}_{\mathrm{fit}} = \mathrm{cov}(\mathbf{X}, \mathbf{f})\mathrm{cov}(\mathbf{X}, \mathbf{f})^T$, where we have assumed $\mathrm{var}(\mathbf{f}_Y) = \mathbf{I}_r$ without loss of generality. Using model (1) for $\mathbf{X}$ we have, $\mathrm{cov}(\mathbf{X}, \mathbf{f}) = \boldsymbol{\Gamma}\mathbf{C}$, where $\mathbf{C} = \mathrm{cov}(\boldsymbol{\nu}_Y, \mathbf{f}_Y)$. Consequently, $\widehat{\boldsymbol{\Sigma}}^{-1}\widehat{\boldsymbol{\Sigma}}_{\mathrm{fit}}$ converges at $\sqrt{n}$ rate to $\boldsymbol{\Sigma}^{-1}\boldsymbol{\Sigma}_{\mathrm{fit}} = (\boldsymbol{\Gamma}\mathbf{V}\boldsymbol{\Gamma}^T + \boldsymbol{\Delta})^{-1}\boldsymbol{\Gamma}\mathbf{C}\mathbf{C}^T\boldsymbol{\Gamma}^{-1}$, and the first $d$ eigenvectors of $\widehat{\boldsymbol{\Sigma}}^{-1}\widehat{\boldsymbol{\Sigma}}_{\mathrm{fit}}$ converge at the $\sqrt{n}$ rate to corresponding eigenvectors of $\boldsymbol{\Sigma}^{-1}\boldsymbol{\Sigma}_{\mathrm{fit}}$.

Using the form $\boldsymbol{\Sigma}^{-1} = \boldsymbol{\Delta}^{-1} - \boldsymbol{\Delta}^{-1}\boldsymbol{\Gamma}(\mathbf{V}^{-1} + \boldsymbol{\Gamma}^T \cdot \boldsymbol{\Delta}^{-1}\boldsymbol{\Gamma})^{-1}\boldsymbol{\Gamma}^T\boldsymbol{\Delta}^{-1}$ and simplifying we find $\boldsymbol{\Sigma}^{-1}\boldsymbol{\Sigma}_{\mathrm{fit}} = \boldsymbol{\Delta}^{-1}\boldsymbol{\Gamma}\mathbf{K}\mathbf{C}\mathbf{C}^T\boldsymbol{\Gamma}^T$, where $\mathbf{K} = (\mathbf{V}^{-1} + \boldsymbol{\Gamma}^T\boldsymbol{\Delta}^{-1}\boldsymbol{\Gamma})^{-1}\mathbf{V}^{-1}$ is a full rank $d \times d$ matrix. Clearly, $\mathrm{span}(\boldsymbol{\Sigma}^{-1}\boldsymbol{\Sigma}_{\mathrm{fit}}) \subseteq \boldsymbol{\Delta}^{-1}\mathcal{S}_{\boldsymbol{\Gamma}}$ with equality if and only if the rank of $\boldsymbol{\Gamma}\mathbf{K}\mathbf{C} \cdot \mathbf{C}^T\boldsymbol{\Gamma}^T$ is equal to $d$. Since $\boldsymbol{\Gamma}$ has full column rank and $\mathbf{K}$ is nonsingular, the rank of $\boldsymbol{\Gamma}\mathbf{K}\mathbf{C}\mathbf{C}^T\boldsymbol{\Gamma}^T$ is equal to $d$ if and only if the rank of $\mathbf{C}\mathbf{C}^T$ is equal to $d$. The result follows since $\boldsymbol{\rho} = \mathbf{V}^{-1/2}\mathbf{C}$, recalling that we have assumed $\mathrm{var}(\mathbf{f}_Y) = \mathbf{I}_r$. $\quad\square$

PROOF OF LEMMA 6.1. $Y \perp\!\!\!\perp \mathbf{X}_2 | \mathbf{X}_1$ if and only if $Y \perp\!\!\!\perp \mathbf{X} | \mathbf{X}_1$. Suppose that $Y \perp\!\!\!\perp \mathbf{X} | \mathbf{X}_1$. We know that $\boldsymbol{\Gamma}^T\boldsymbol{\Delta}^{-1}\mathbf{X}$ is the minimal sufficient reduction and thus it should not depend on $\mathbf{X}_2$. Now,

$$(15) \quad \begin{aligned} \boldsymbol{\Gamma}^T\boldsymbol{\Delta}^{-1}\mathbf{X} &= (\boldsymbol{\Gamma}_1^T\boldsymbol{\Delta}_{11} + \boldsymbol{\Gamma}_2^T\boldsymbol{\Delta}_{21})\mathbf{X}_1 \\ &\quad + (\boldsymbol{\Gamma}_1^T\boldsymbol{\Delta}_{12} + \boldsymbol{\Gamma}_2^T\boldsymbol{\Delta}_{22})\mathbf{X}_2 \end{aligned}$$

will not depend on $\mathbf{X}_2$ if and only if $\boldsymbol{\Gamma}_1^T\boldsymbol{\Delta}^{12} + \boldsymbol{\Gamma}_2^T\boldsymbol{\Delta}^{22} = 0$ equivalently $\boldsymbol{\Gamma}_2 = -\boldsymbol{\Delta}^{-22}\boldsymbol{\Delta}^{21}\boldsymbol{\Gamma}_1$. The reciprocal follows directly if we replace $\boldsymbol{\Gamma}_2$ by $-\boldsymbol{\Delta}^{-22}\boldsymbol{\Delta}^{21}\boldsymbol{\Gamma}_1$ on equation (15). $\quad\square$

PROOF OF THEOREM 6.2. After maximizing the log likelihood over $(\boldsymbol{\mu}, \boldsymbol{\beta})$, we need to maximize on $\boldsymbol{\Gamma}$ and $\boldsymbol{\Delta}^{-1}$ $f(\boldsymbol{\Gamma}, \boldsymbol{\Delta}^{-1}) = \log|\boldsymbol{\Delta}^{-1}| - \mathrm{tr}(\boldsymbol{\Delta}^{-1}\widehat{\boldsymbol{\Sigma}}) + \mathrm{tr}(\boldsymbol{\Delta}^{-1}\mathbf{P}_{\boldsymbol{\Gamma}(\boldsymbol{\Delta}^{-1})}\widehat{\boldsymbol{\Sigma}}_{\mathrm{fit}})$. From the hypotheses on $\boldsymbol{\Gamma}$, we have $\boldsymbol{\Gamma}^T\boldsymbol{\Delta}^{-1} = (\boldsymbol{\Gamma}_1^T\boldsymbol{\Delta}^{11.2}, 0)$ where $\boldsymbol{\Delta}^{11.2} = \boldsymbol{\Delta}^{11} - \boldsymbol{\Delta}^{12}\boldsymbol{\Delta}^{-22}\boldsymbol{\Delta}^{21}$. Then $\boldsymbol{\Gamma}^T\boldsymbol{\Delta}^{-1}\boldsymbol{\Gamma} = \boldsymbol{\Gamma}_1^T\boldsymbol{\Delta}^{11.2}\boldsymbol{\Gamma}_1$. For fixed $\boldsymbol{\Delta}$, the last term is maximized by choosing $(\boldsymbol{\Delta}^{11.2})^{1/2} \cdot \boldsymbol{\Gamma}_1$ to be a basis for the span the first $d$ eigenvectors of $(\boldsymbol{\Delta}^{11.2})^{1/2} \widehat{\boldsymbol{\Sigma}}_{11,\mathrm{fit}} (\boldsymbol{\Delta}^{11.2})^{1/2}$, yielding another partially maximized log likelihood

$$(16) \quad f(\boldsymbol{\Delta}^{-1}) = \log |\boldsymbol{\Delta}^{-1}| - \mathrm{tr}(\boldsymbol{\Delta}^{-1} \widehat{\boldsymbol{\Sigma}}) \\ + \sum_{i=1}^{d} \lambda_i \{ (\boldsymbol{\Delta}^{11.2})^{1/2} \widehat{\boldsymbol{\Sigma}}_{11,\mathrm{fit}} (\boldsymbol{\Delta}^{11.2})^{1/2} \}.$$

Let us take the one-to-one and onto transformation defined by $\mathbf{L}_{11} = \boldsymbol{\Delta}^{11} - \boldsymbol{\Delta}^{12} \boldsymbol{\Delta}^{-22} \boldsymbol{\Delta}^{21}$, $\mathbf{L}_{22} = \boldsymbol{\Delta}^{22}$ and $\mathbf{L}_{12} = \boldsymbol{\Delta}^{12} \boldsymbol{\Delta}^{-22}$. As a function of $\mathbf{L}_{11}$, $\mathbf{L}_{22}$, $\mathbf{L}_{12}$ we get

$$f(\mathbf{L}_{11}, \mathbf{L}_{22}, \mathbf{L}_{12}) \\ = \log |\mathbf{L}_{11}| + \log |\mathbf{L}_{22}| \\ - \mathrm{tr}(\mathbf{L}_{22} \widehat{\boldsymbol{\Sigma}}_{22} + \mathbf{L}_{22} \mathbf{L}_{12}^T \widehat{\boldsymbol{\Sigma}}_{12}) \\ - \mathrm{tr}\{ (\mathbf{L}_{11} + \mathbf{L}_{12} \mathbf{L}_{22} \mathbf{L}_{12}^T) \widehat{\boldsymbol{\Sigma}}_{11} \\ + \mathbf{L}_{12} \mathbf{L}_{22} \widehat{\boldsymbol{\Sigma}}_{21} \} \\ + \sum_{i=1}^{d} \lambda_i (\mathbf{L}_{11}^{1/2} \widehat{\boldsymbol{\Sigma}}_{11,\mathrm{fit}} \mathbf{L}_{11}^{1/2}).$$

Now, differentiating with respect to $\mathbf{L}_{12}$ in the last expression, we get that

$$\frac{\partial f}{\partial \mathbf{L}_{12}} = -2 \widehat{\boldsymbol{\Sigma}}_{12} \mathbf{L}_{22} - 2 \widehat{\boldsymbol{\Sigma}}_{11} \mathbf{L}_{12} \mathbf{L}_{22} \quad \text{and}$$

$$\frac{\partial^2 f}{\partial \mathbf{L}_{12}^2} = -2 \widehat{\boldsymbol{\Sigma}}_{11} \otimes \mathbf{L}_{22}.$$

Therefore the maximum occurs when $\mathbf{L}_{12} = -\widehat{\boldsymbol{\Sigma}}_{11}^{-1} \cdot \widehat{\boldsymbol{\Sigma}}_{12}$. Replacing this in the last log likelihood function we next need to maximize

$$f(\mathbf{L}_{11}, \mathbf{L}_{22}) = \log |\mathbf{L}_{11}| + \log |\mathbf{L}_{22}| \\ - \mathrm{tr}(\mathbf{L}_{22} \widehat{\boldsymbol{\Sigma}}_{22}) - \mathrm{tr}(\mathbf{L}_{11} \widehat{\boldsymbol{\Sigma}}_{11}) \\ + \mathrm{tr}(\mathbf{L}_{22} \widehat{\boldsymbol{\Sigma}}_{12} \widehat{\boldsymbol{\Sigma}}_{11}^{-1} \widehat{\boldsymbol{\Sigma}}_{12}) \\ + \sum_{i=1}^{d} \lambda_i (\mathbf{L}_{11}^{1/2} \widehat{\boldsymbol{\Sigma}}_{11,\mathrm{fit}} \mathbf{L}_{11}^{1/2}),$$

since for $\mathbf{L}_{12} = -\widehat{\boldsymbol{\Sigma}}_{11}^{-1} \widehat{\boldsymbol{\Sigma}}_{12}$, $-2 \mathrm{tr}(\mathbf{L}_{22} \mathbf{L}_{12}^T \widehat{\boldsymbol{\Sigma}}_{12}) - \mathrm{tr}(\mathbf{L}_{12} \mathbf{L}_{22} \mathbf{L}_{12}^T \widehat{\boldsymbol{\Sigma}}_{11}) = \mathrm{tr}(\mathbf{L}_{22} \widehat{\boldsymbol{\Sigma}}_{12} \widehat{\boldsymbol{\Sigma}}_{11}^{-1} \widehat{\boldsymbol{\Sigma}}_{12})$. The maximum on $\mathbf{L}_{22}$ is at $\mathbf{L}_{22} = \widehat{\boldsymbol{\Sigma}}_{22.1}^{-1}$ so that we need to maximize on $\mathbf{L}_{11}$

$$f(\mathbf{L}_{11}) = \log |\mathbf{L}_{11}| - \mathrm{tr}(\mathbf{L}_{11} \widehat{\boldsymbol{\Sigma}}_{11}) \\ + \sum_{i=1}^{d} \lambda_i (\mathbf{L}_{11}^{1/2} \widehat{\boldsymbol{\Sigma}}_{11,\mathrm{fit}} \mathbf{L}_{11}^{1/2}).$$

From Theorem 3.1 the MLE for $\mathbf{L}_{11}^{-1}$ is $\widehat{\boldsymbol{\Sigma}}_{11,\mathrm{res}}^{1/2} \widehat{\mathbf{V}} (\mathbf{I}_d + \widehat{\mathbf{K}}) \widehat{\mathbf{V}}^T \widehat{\boldsymbol{\Sigma}}_{11,\mathrm{res}}^{1/2}$, where $\widehat{\mathbf{K}}$ and $\widehat{\mathbf{V}}$ are as defined in Theorem 6.2. Since $\mathbf{L}_{11} = \boldsymbol{\Delta}^{11} - \boldsymbol{\Delta}^{12} \boldsymbol{\Delta}^{-22} \boldsymbol{\Delta}^{21} = \boldsymbol{\Delta}_{11}^{-1}$ it follows that $\widehat{\boldsymbol{\Delta}}_{11} = \widehat{\boldsymbol{\Sigma}}_{11,\mathrm{res}}^{1/2} \widehat{\mathbf{V}} (\mathbf{I}_d + \widehat{\mathbf{K}}) \widehat{\mathbf{V}}^T \widehat{\boldsymbol{\Sigma}}_{11,\mathrm{res}}^{1/2}$. The MLE for $\boldsymbol{\Delta}^{22}$ is $\widehat{\boldsymbol{\Sigma}}_{22.1}^{-1}$ and for $\boldsymbol{\Delta}^{12}$ is $-\widehat{\boldsymbol{\Sigma}}_{11}^{-1} \widehat{\boldsymbol{\Sigma}}_{12} \widehat{\boldsymbol{\Sigma}}_{22.1}^{-1}$. Consequently, $\widehat{\boldsymbol{\Delta}}_{12} = \widehat{\boldsymbol{\Delta}}_{11} \widehat{\boldsymbol{\Sigma}}_{11}^{-1} \widehat{\boldsymbol{\Sigma}}_{12}$ and $\widehat{\boldsymbol{\Delta}}_{22} = \widehat{\boldsymbol{\Sigma}}_{22.1} + \widehat{\boldsymbol{\Sigma}}_{21} \widehat{\boldsymbol{\Sigma}}_{11}^{-1} \widehat{\boldsymbol{\Delta}}_{11} \widehat{\boldsymbol{\Sigma}}_{11}^{-1} \widehat{\boldsymbol{\Sigma}}_{12}$.

The MLE for the span of $\boldsymbol{\Delta}^{-1} \boldsymbol{\Gamma} = (\boldsymbol{\Delta}^{11.2} \boldsymbol{\Gamma}_1, 0)^T$ is the span of $(\widehat{\boldsymbol{\Delta}}_{11}^{-1/2} \widehat{\boldsymbol{\Gamma}}_1, 0)^T$ with $\widehat{\boldsymbol{\Gamma}}_1$ the first $d$ eigenvectors of $\widehat{\boldsymbol{\Delta}}_{11}^{-1/2} \widehat{\boldsymbol{\Sigma}}_{11,\mathrm{fit}} \widehat{\boldsymbol{\Delta}}_{11}^{-1/2}$. Using the logic of Corollary 3.3 it can be proved that the MLE of span$(\boldsymbol{\Delta}^{-1} \boldsymbol{\Gamma})$ is in this case the span of $(\widehat{\boldsymbol{\Sigma}}_{11,\mathrm{res}}^{-1/2} \widehat{\boldsymbol{\Gamma}}_1, 0)^T$, with $\widehat{\boldsymbol{\Gamma}}_1$ the first $d$ eigenvectors of $\widehat{\boldsymbol{\Sigma}}_{11,\mathrm{res}}^{-1/2} \widehat{\boldsymbol{\Sigma}}_{11,\mathrm{fit}} \widehat{\boldsymbol{\Sigma}}_{11,\mathrm{res}}^{-1/2}$.

The proof of (6) can be done essentially as the proof of (5). $\square$

## APPENDIX B: ALGORITHM FOR $\boldsymbol{\Delta}$ WITH LINEAR STRUCTURE

We will maximize (4) as a function $f$ of $\mathbf{S} = \boldsymbol{\Delta}^{-1}$. We first find the derivative with respect to $\mathbf{S}$ without considering any structure. Because $\widehat{\boldsymbol{\Sigma}}_{\mathrm{fit}}$ is symmetric we get

$$\frac{\partial f(\mathbf{S})}{\partial \mathbf{S}} = \mathbf{S}^{-1} - \widehat{\boldsymbol{\Sigma}}_{\mathrm{res}} - \sum_{i=d+1}^{r} \mathbf{v}_i (\mathbf{S} \widehat{\boldsymbol{\Sigma}}_{\mathrm{fit}}) \mathbf{u}_i^T (\mathbf{S} \widehat{\boldsymbol{\Sigma}}_{\mathrm{fit}}) \widehat{\boldsymbol{\Sigma}}_{\mathrm{fit}},$$

where $\mathbf{u}_i$ and $\mathbf{v}_i$ indicate respectively the normalized right and left eigenvectors corresponding to the eigenvalue $\lambda_i$ of $\mathbf{S} \widehat{\boldsymbol{\Sigma}}_{\mathrm{fit}}$. Now, $\partial \mathbf{S} / \partial s_h = \mathbf{G}_h$ and

$$(17) \quad \frac{\partial f(\mathbf{S})}{\partial s_h} = \mathrm{tr}(\mathbf{S}^{-1} \mathbf{G}_h) - \mathrm{tr}(\widehat{\boldsymbol{\Sigma}}_{\mathrm{res}} \mathbf{G}_h) \\ - \sum_{i=d+1}^{r} \mathrm{tr}\{ \widehat{\boldsymbol{\Sigma}}_{\mathrm{fit}} \mathbf{u}_i (\mathbf{S} \widehat{\boldsymbol{\Sigma}}_{\mathrm{fit}}) \mathbf{v}_i^T (\mathbf{S} \widehat{\boldsymbol{\Sigma}}_{\mathrm{fit}}) \mathbf{G}_h \}.$$

Denote by $\bar{\mathbf{u}}_i$ the $i$th eigenvector of $\mathbf{S}^{1/2} \widehat{\boldsymbol{\Sigma}}_{\mathrm{fit}} \mathbf{S}^{1/2} = \boldsymbol{\Delta}^{-1/2} \widehat{\boldsymbol{\Sigma}}_{\mathrm{fit}} \boldsymbol{\Delta}^{-1/2}$ corresponding to the $\lambda_i$ eigenvalue (in decreasing order) normalized with unit norm. Then $\mathbf{u}_i = \mathbf{S}^{1/2} \bar{\mathbf{u}}_i = \mathbf{D}^{-1/2} \bar{\mathbf{u}}_i$, $\mathbf{v}_i = \mathbf{S}^{-1/2} \bar{\mathbf{u}}_i = \mathbf{D}^{1/2} \bar{\mathbf{u}}_i$, $\widehat{\boldsymbol{\Sigma}}_{\mathrm{fit}} \mathbf{u}_i = \lambda_i \mathbf{D}^{1/2} \bar{\mathbf{u}}_i$ and $\mathbf{v}_i^T \mathbf{u}_j = 0$ if $i \neq j$ and 1 otherwise. We can rewrite (17) as

$$(18) \quad \frac{\partial f(\mathbf{S})}{\partial s_h} = \mathrm{tr}(\boldsymbol{\Delta} \mathbf{G}_h) - \mathrm{tr}(\widehat{\boldsymbol{\Sigma}}_{\mathrm{res}} \mathbf{G}_h) \\ - \sum_{i=d+1}^{r} \lambda_i \mathrm{tr}(\boldsymbol{\Delta}^{1/2} \bar{\mathbf{u}}_i \bar{\mathbf{u}}_i^T \boldsymbol{\Delta}^{1/2} \mathbf{G}_h).$$

To find the MLE we need to solve $\partial f(\mathbf{S}) / \partial s_h = 0$ for $h = 1, \ldots, m$. Using (18) we can rewrite $\partial f(\mathbf{S}) / \partial s_h =$

0 using the vec operator as

$$
\begin{aligned}
\text{(19)} \quad & \text{vec}(\mathbf{G}_h)^T \text{vec}(\boldsymbol{\Delta}) \\
& = \text{vec}(\mathbf{G}_h)^T \text{vec}(\widehat{\boldsymbol{\Sigma}}_{\text{res}}) \\
& \quad + \text{vec}(\mathbf{G}_h)^T \sum_{i=d+1}^{r} \lambda_i \text{vec}(\boldsymbol{\Delta}^{1/2} \bar{\mathbf{u}}_i \bar{\mathbf{u}}_i^T \boldsymbol{\Delta}^{1/2})
\end{aligned}
$$

for $h = 1, \dots, m$. Let $\tilde{\mathbf{G}} = \{\text{vec}(\mathbf{G}_1), \dots, \text{vec}(\mathbf{G}_m)\}$. Since $\boldsymbol{\Delta} = \sum_{i=1}^{m} \delta_i \mathbf{G}_i$, $\text{vec}(\boldsymbol{\Delta}) = \tilde{\mathbf{G}} \boldsymbol{\delta}$ we can rewrite (19) for all $h$ as

$$
\begin{aligned}
\text{(20)} \quad \tilde{\mathbf{G}}^T \tilde{\mathbf{G}} \boldsymbol{\delta} = \tilde{\mathbf{G}}^T \bigg\{ & \text{vec}(\widehat{\boldsymbol{\Sigma}}_{\text{res}}) \\
& + \sum_{i=d+1}^{r} \lambda_i \text{vec}(\boldsymbol{\Delta}^{1/2} \bar{\mathbf{u}}_i \bar{\mathbf{u}}_i^T \boldsymbol{\Delta}^{1/2}) \bigg\}.
\end{aligned}
$$

Now, if $r = d$ we get $\boldsymbol{\delta} = (\tilde{\mathbf{G}}^T \tilde{\mathbf{G}})^{-1} \tilde{\mathbf{G}}^T \text{vec}(\widehat{\boldsymbol{\Sigma}}_{\text{res}})$ and $\boldsymbol{\Delta}^{-1} = (\sum_{i=1}^{m} \delta_i \mathbf{G}_i)^{-1}$. The algorithm will be:

1. Set $\boldsymbol{\delta}_0 = (\delta_1^0, \dots, \delta_m^0) = (\tilde{\mathbf{G}}^T \tilde{\mathbf{G}})^{-1} \tilde{\mathbf{G}}^T \text{vec}(\widehat{\boldsymbol{\Sigma}}_{\text{res}})$.
2. Compute $\boldsymbol{\Delta}_0 = \sum_{i=1}^{m} \delta_i^0 \mathbf{G}_i$ and $\mathbf{S}_0 = \boldsymbol{\Delta}_0^{-1}$.
3. Compute until convergence, $n = 1, 2, \dots$,

$$
\begin{aligned}
\boldsymbol{\Delta}_n = (\tilde{\mathbf{G}}^T \tilde{\mathbf{G}})^{-1} \tilde{\mathbf{G}}^T \\
\cdot \bigg[ & \text{vec}(\widehat{\boldsymbol{\Sigma}}_{\text{res}}) \\
& + \sum_{i=d+1}^{r} \lambda_i^{\mathbf{S}_{n-1}} \text{vec}\{\boldsymbol{\Delta}_{n-1}^{1/2} \bar{\mathbf{u}}_i^{\mathbf{S}_{n-1}} (\bar{\mathbf{u}}_i^{\mathbf{S}_{n-1}})^T \boldsymbol{\Delta}_{n-1}^{1/2}\} \bigg]
\end{aligned}
$$

with $\mathbf{S}_{n-1} = \boldsymbol{\Delta}_{n-1}^{-1}$ and $\lambda^{\mathbf{S}_{n-1}}$ and $\bar{u}^{\mathbf{S}_{n-1}}$ denoting respectively the eigenvalues and eigenvectors of $\mathbf{S}_{n-1}^{1/2} \widehat{\boldsymbol{\Sigma}}_{\text{fit}} \mathbf{S}_{n-1}^{1/2}$.

## ACKNOWLEDGMENTS

The authors thank Prof. Eaton for his helpful discussions. Research for this article was supported in part by NSF Grant DMS-07-04098.